\documentclass[pdflatex,referee,sn-nature]{sn-jnl}

\usepackage{graphicx}%
\usepackage{multirow}%
\usepackage{amsmath,amssymb,amsfonts}%
\usepackage{amsthm}%
\usepackage{mathrsfs}%
\usepackage[title]{appendix}%
\usepackage{xcolor}%
\usepackage{textcomp}%
\usepackage{manyfoot}%
\usepackage{booktabs}%
\usepackage{algorithm}%
\usepackage{algorithmicx}%
\usepackage{algpseudocode}%
\usepackage{listings}%
\usepackage{setspace}%

\def\aj{{AJ}}                   
\def\araa{{ARA\&A}}          
\def\apj{{ApJ}}                 
\def\apjl{{ApJ}}                
\def\apjs{ {ApJS}}

\def\aap{ {A\&A}}

\def\mnras{ {MNRAS}}

\def\pasj{ {PASJ}}

\def\nat{ {Nature}}

\def\physrep{ {Phys.~Rep.}}

\begin{document}

\title[A Fast, Hot Wind from a Nuclear Starburst]{A Fast, Hot Wind from a Nuclear Starburst}


\author[1]{\textit{XRISM} Collaboration}
\author[2]{Marc Audard}
\author[3]{Hisamitsu Awaki}
\author[4,5,6]{Ralf Ballhausen}
\author[7]{Aya Bamba}
\author[8]{Ehud Behar}
\author[9,5,6]{Rozenn Boissay-Malaquin}
\author[10]{Laura Brenneman}
\author[11]{Gregory V.\ Brown}
\author[12]{Lia Corrales}
% ...

\affil[1]{Corresponding Author: Erin Boettcher (eboettch@umd.edu)}
\affil[2]{Department of Astronomy, University of Geneva, Versoix CH-1290, Switzerland} 
\affil[3]{Department of Physics, Ehime University, Ehime 790-8577, Japan} 
\affil[4]{Department of Astronomy, University of Maryland, College Park, MD 20742, USA} 
\affil[5]{NASA / Goddard Space Flight Center, Greenbelt, MD 20771, USA}
\affil[6]{Center for Research and Exploration in Space Science and Technology, NASA / GSFC (CRESST II), Greenbelt, MD 20771, USA}
\affil[7]{Department of Physics, University of Tokyo, Tokyo 113-0033, Japan} 
\affil[8]{Department of Physics, Technion, Technion City, Haifa 3200003, Israel} 
\affil[9]{Center for Space Sciences and Technology, University of Maryland, Baltimore County (UMBC), Baltimore, MD, 21250 USA}
\affil[10]{Center for Astrophysics | Harvard-Smithsonian, MA 02138, USA} 
\affil[11]{Lawrence Livermore National Laboratory, CA 94550, USA} 
\affil[12]{Department of Astronomy, University of Michigan, MI 48109, USA} 
% ...

\maketitle

\clearpage

\textbf{Galaxies with intense star formation often host multiphase, galaxy-scale winds powered by supernovae and fast stellar winds \cite{1978ApJ...219L..23M, 1996ApJ...462..651L, 1990ApJS...74..833H}. These are strong enough to disrupt the star-forming interstellar medium, and they chemically enrich the surrounding circumgalactic medium \cite{2001ApJ...557..605S, 2015ARA&A..53...51S, 2023ARA&A..61..131F}. However, their launching mechanism remains unknown \cite{2024ARA&A..62..529T}. Here we show that thermal gas pressure is sufficient to drive the multiphase wind in the prototypical starburst galaxy M82. Using a high energy-resolution ($\Delta E = 4.5$ eV) \textit{XRISM} Resolve spectrum, including detections of Fe~{\sc xxv}~6.7~keV, Ar~{\sc xvii}~3.1~keV, and S~{\sc xvi}~2.6 keV, we measure the temperature ($T = 2.3^{+0.5}_{-0.2} \times 10^7$~K) and mass ($M \approx 6 \pm 2 \times 10^5$~M$_\odot$) of the hot gas in the starburst and provide the first direct measurement of its line-of-sight velocity dispersion ($\sigma = 595^{+464}_{-128}$~km~s$^{-1}$). These values are consistent with a freely-expanding wind exceeding the galactic escape velocity. The size of the Fe~{\sc xxv}-emitting region suggests a hot gas outflow rate of $\dot{M} \approx 4$~M$_\odot$~yr$^{-1}$, carrying a total energy of $\dot{E} \approx 2 \times 10^{42}$~erg~s$^{-1}$. This is sufficient to drive the molecular, atomic, and ionized outflows while transporting up to $\approx 2$~M$_\odot$~yr$^{-1}$ of hot gas to the intergalactic medium. The estimated supernova rate implies that $\approx 60$\% of the supernova energy must be thermalized in hot gas. Our results suggest that additional driving mechanisms, such as cosmic-ray pressure, are not required to launch the wind.} \\


The galaxy M82 ($D \approx 3.5$~Mpc \cite{1994ApJ...427..628F}) hosts a spectacular, kiloparsec-scale wind \cite{2009ApJ...697.2030S} that transports at least $30$~M$_{\odot}$~yr$^{-1}$ of molecular \cite{2002ApJ...580L..21W, 2009ApJ...700L.149V, 2015ApJ...814...83L}, neutral \cite{2018ApJ...856...61M}, and warm ionized gas \cite{1987AJ.....93..264M, 1998ApJ...493..129S} out of the galaxy. The outflow is driven by a population of nuclear star clusters ($R \lesssim 500$ pc) produced by bursty star formation over the past $\approx 15$~Myr \cite{2003ApJ...599..193F}. A nuclear starburst event that was driven by dynamical torques from M81 peaked $\approx 10$ Myr ago \cite{2003ApJ...599..193F} and cleared dense interstellar gas from the inner galaxy \cite{1987PASJ...39..685N, 2001A&A...365..571W}, enabling a galaxy-scale wind to break out of the disk \cite{1999ApJ...523..575L, 2009ApJ...697.2030S}. A second burst of star formation occurred in a circumnuclear ring and along the stellar bar $\approx 5$ Myr ago \cite{2003ApJ...599..193F}. Stellar winds and supernovae (SNe) from these bursts, as well as from ongoing star formation at a rate of $\text{SFR} \approx 10$~M$_{\odot}$~yr$^{-1}$ \cite{2018ApJ...864..150V}, shock-heat the ambient interstellar medium (ISM) to $T = 10^{7} - 10^{8}$ K. This produces a hot wind fluid seen directly in diffuse, hard X-ray emission in the inner few hundred parsecs of the galaxy \cite{2007ApJ...658..258S, 2021A&A...652A..18I}. Over the past forty years, analytic models \cite{1985Natur.317...44C} have posited that the thermal pressure of this hot wind fluid is responsible for driving the kiloparsec-scale winds that are common around highly star-forming galaxies (e.g., \cite{1990ApJS...74..833H, 1996ApJ...462..651L}). However, existing measurements are unable to confirm this. Here we use measurements from the nucleus of M82 from the Resolve microcalorimeter on the \textit{X-ray Imaging and Spectroscopy Mission} (\textit{XRISM}) to determine the total energy carried by the hot wind fluid and assess whether gas pressure can drive the hot and cool winds.

\begin{figure}[]
\centering
\includegraphics[scale = 0.22]{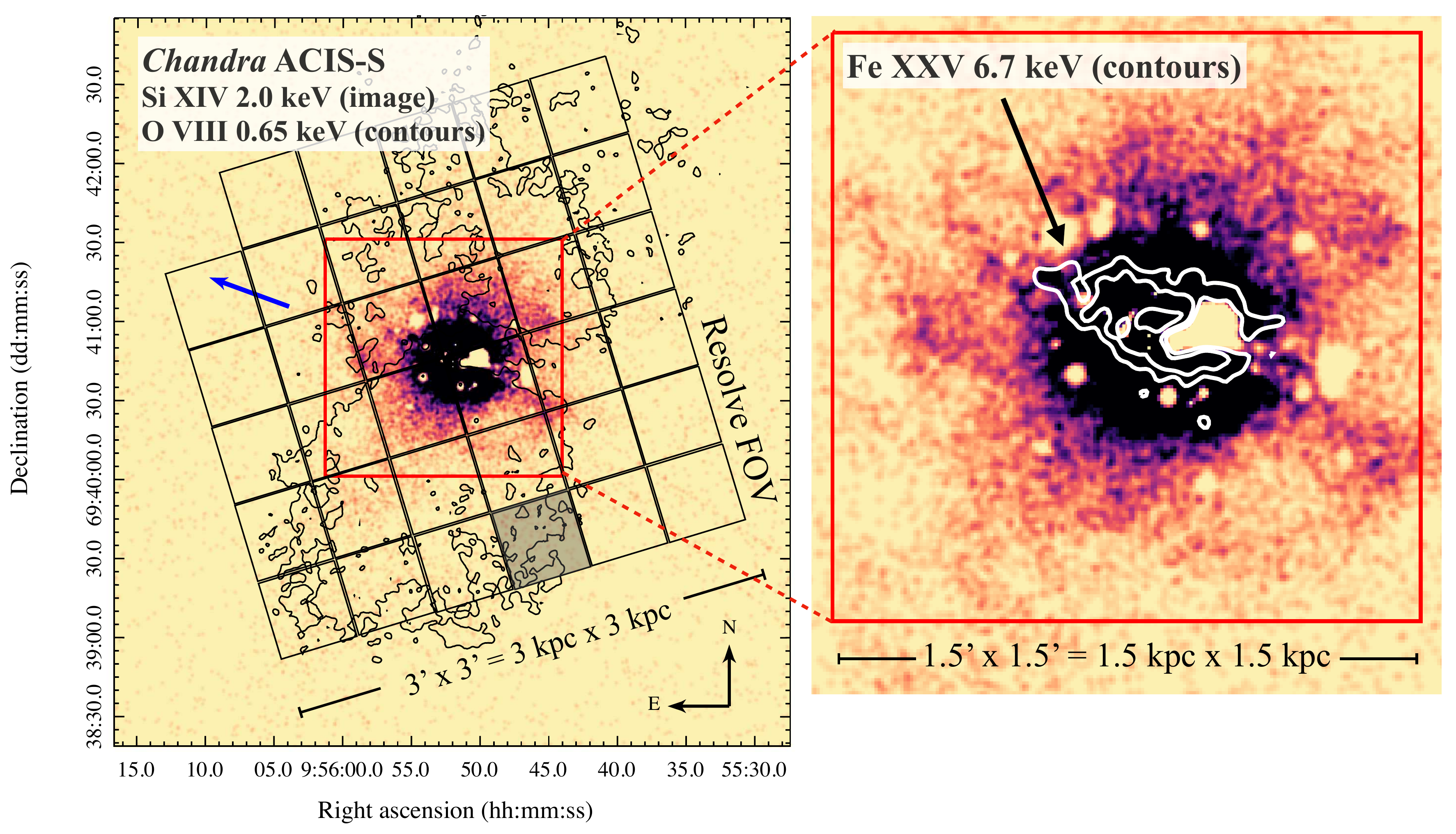}
\caption{\textbf{The hot starburst wind in M82.} The soft X-ray wind is seen at left in a \textit{Chandra} ACIS-S narrowband image at Si~{\sc xiv}~2.0~keV and contour plot of O~{\sc viii}~0.65~keV emission. \textit{Chandra} reveals that the spatial extent of Si~{\sc xiv} emission ($R \lesssim 1$~kpc) is representative of the $E < 4$~keV emission lines seen by Resolve, including S~{\sc xv}~2.4~keV, S~{\sc xvi}~2.6~keV, and Ar~{\sc xvii}~3.1~keV. Softer X-ray emission is extended to $R \gtrsim 2$~kpc, as shown by the extent of the O~{\sc viii} emission. As shown at right, Fe~{\sc xxv}~6.7~keV emission is largely contained within a projected ellipsoid with $a \approx 300$~pc, $c \approx 75$~pc (white contours). This suggests that the Fe~{\sc xxv} emission largely traces the thermalization zone, while the $E < 4$~keV lines are produced in and around the starburst nucleus and in the biconical outflow. The nucleus falls fully within the Resolve field of view ($3' \times 3' = 3~\text{kpc} \times 3~\text{kpc}$; black outline at left, with the excluded pixel 27 shaded in gray). Point sources have been masked in both images. The blue arrow lies along the major axis of the galaxy.}
\label{pointing_fig}
\end{figure}

\begin{figure}[]
\centering
\includegraphics[scale = 0.35]{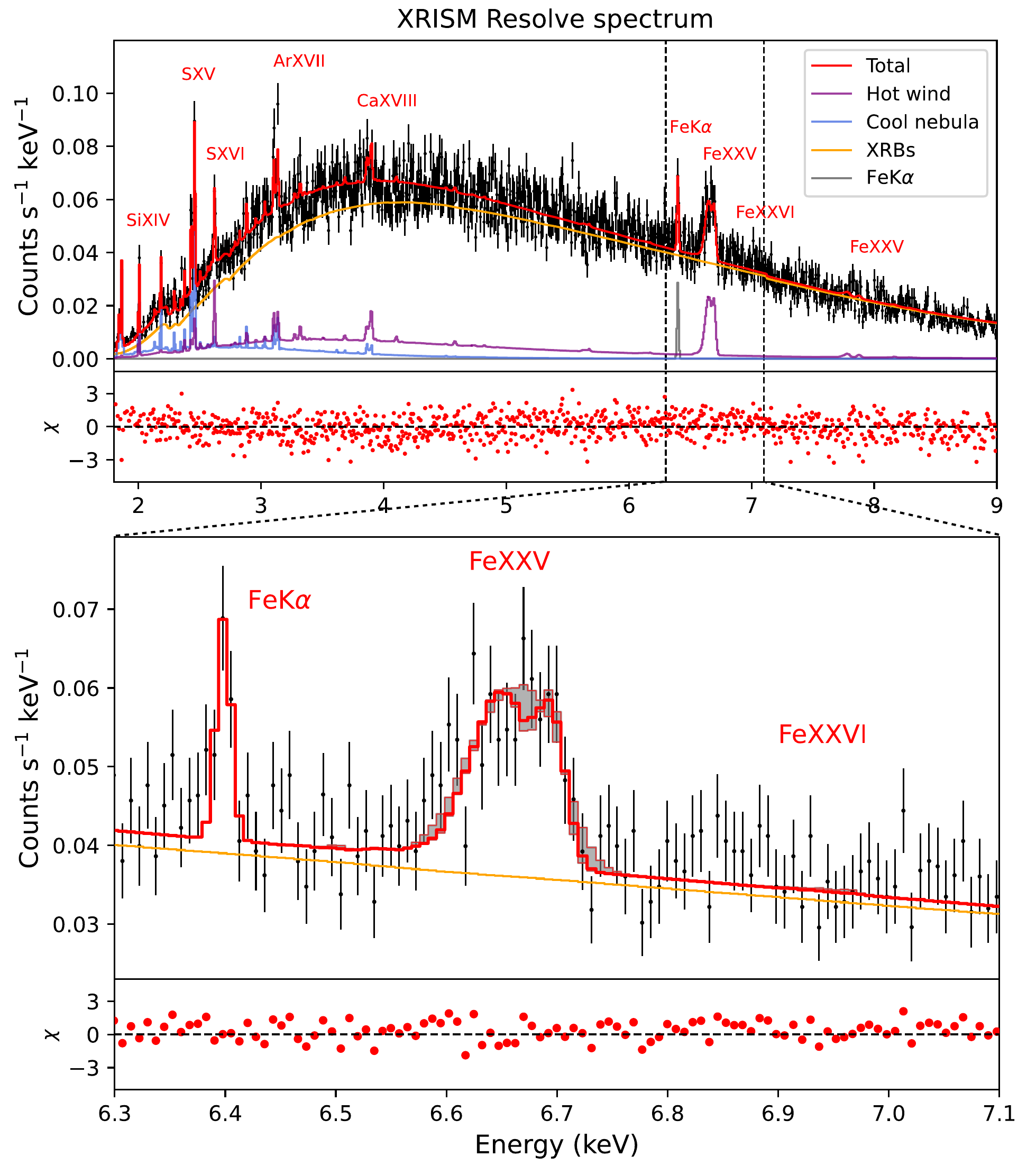}
\caption{\textbf{Resolve spectrum of the starburst nucleus of M82.} At top, the Resolve spectrum (black histogram) shows a suite of resolved emission lines, including strong Fe~{\sc xxv}~6.7~keV emission. The best-fit model (red) consists of a hot, $kT = 2.0$~keV wind fluid (purple), a cooler, $kT = 0.7$~keV phase, a powerlaw continuum from X-ray binaries (XRBs; orange), and a Gaussian Fe~K$\alpha$~6.4~keV emission line (gray). This fluorescence line arises from neutral gas irradiated by hard X-ray photons from compact objects \cite{2007ApJ...658..258S, 2014MNRAS.437L..76L}. In the bottom panel, the significant velocity broadening of the $kT = 2$~keV gas ($\sigma = 595^{+464}_{-128}$~km~s$^{-1}$) is evident from the visible blending of the Fe~{\sc xxv} triplet lines. The gray shading shows the suite of best-fit models obtained by varying $\sigma$ within the uncertainty.}
\label{full_spec}
\end{figure}

Archival \textit{Chandra X-ray Observatory} data reveals the spatial distribution of hot gas of different temperatures as traced by characteristic strong emission lines in and around the starburst nucleus of M82 (Figure~\ref{pointing_fig}). Most of the Fe~{\sc xxv}~6.7~keV emission, which traces the hottest gas, falls within a projected ellipsoid with $a \approx 300$~pc, $c \approx 75$~pc \cite{2007ApJ...658..258S} that is surrounded by a more extended nebula ($R \approx 1$~kpc) that is bright in the softer lines that trace cooler gas ($E \lesssim 3$~keV). We observed this region of M82 with \textit{XRISM} Resolve \cite{2020SPIE11444E..22T} ($E = 1.8 - 10$~keV; $\Delta E = 4.5$~eV) (see Methods and Supplementary Information). In Figure~\ref{full_spec}, we show the Resolve spectrum. The key feature of the spectrum is a strong detection of Fe~{\sc xxv}~6.7~keV emission. We also detect a suite of lower energy emission lines, the strongest of which include Si~{\sc xiii}~1.8~keV, Si~{\sc xiv}~2.0~keV, S~{\sc xv}~2.4~keV, S~{\sc xvi}~2.6~keV, and Ar~{\sc xvii}~3.1~keV. With the high energy resolution of Resolve, we can measure velocity centroids, widths, and more accurate fluxes of these emission lines than with CCD data, enabling precise measurements of the temperature and velocity of the hot wind for the first time.

We fit the broad-band spectrum ($E = 1.8 - 10$~keV) with a two-temperature, velocity-broadened thermal plasma model in collisional ionization equilibrium. The broad-band fit maximizes simultaneous constraints on the hot wind fluid from Fe~{\sc xxv}~6.7~keV and intermediate-energy emission lines, including S~{\sc xvi}~2.6~keV, that also arise from the hot phase. The details of the fitting, including the measured properties of the cool phase ($kT = 0.72^{+0.10}_{-0.08}$~keV; $\sigma = 175^{+86}_{-73}$~km~s$^{-1}$), are discussed in the Methods (see also Table~\ref{bf_model_tab}). The more precise line fluxes in the Resolve observation compared to prior measurements using CCD data enable improved constraints on the hot gas temperature, velocity, and normalization from Fe~{\sc xxv}~6.7~keV and lower energy lines (e.g., the temperature-sensitive S~{\sc xvi}/S~{\sc xv} ratio). This yields a hot, fast wind, with temperature $kT = 2.0^{+0.39}_{-0.15}$~keV, or $T = 2.3^{+0.5}_{-0.2} \times 10^{7}$~K, and line-of-sight velocity dispersion $\sigma = 595^{+464}_{-128}$~km~s$^{-1}$. This rules out much of the previously estimated temperature range from CCD spectroscopy, $T \approx 3 - 8 \times 10^{7}$~K \cite{2000Sci...290.1325G, 2009ApJ...697.2030S}. As discussed in the Methods, we also derive an independent, $3\sigma$ upper limit on the gas temperature of $kT < 5.5$~keV, or $T < 6 \times 10^{7}$~K, from the non-detection of Fe~{\sc xxvi}~6.9~keV ($f(\text{Fe~{\sc xxvi}}) < 3 \times 10^{-6}$~ph~s$^{-1}$~cm$^{-2}$) and Fe~{\sc xxv} to Fe~{\sc xxvi} photon ratio ($f(\text{Fe~{\sc xxv}})/f(\text{Fe~{\sc xxvi}}) > 5$). Combined with the global fit, this limit rules out the upper end of the temperature range from CCD measurements at high significance.

The significant velocity broadening, previously unmeasurable with CCDs, is evident from visual inspection of the Fe~{\sc xxv}~6.7~keV line profile (Figure~\ref{full_spec}), which shows intrinsic blending of the He$\alpha$ triplet lines. The disk of M82 is highly inclined with respect to the line of sight ($i \approx 77^{\circ}$ \cite{2005ApJ...628L..33M}), and the hot wind is collimated into a bicone oriented along the minor axis. The outflow velocity is therefore preferentially in the plane of the sky, and the true outflow speed is likely larger than the line-of-sight velocity dispersion. Even without correcting for projection effects, $v_{\text{out}} = \sigma$ is above the galactic escape velocity of $v_{\text{esc}} \lesssim 450$~km~s$^{-1}$ \cite{2009ApJ...697.2030S} (although note that estimating $v_{\text{esc}}$ for M82 is complicated by the interaction with M81). The velocity centroid of the Fe~{\sc xxv} emission, $v = 189^{+250}_{-145}$~km~s$^{-1}$, is consistent within the errors with the systemic velocity ($v_{\text{sys}} \approx 270$~km~s$^{-1}$), as expected for a quasi-symmetric outflow (all velocities are given in the heliocentric frame).

\begin{figure}[]
\centering
\includegraphics[scale = 0.3]{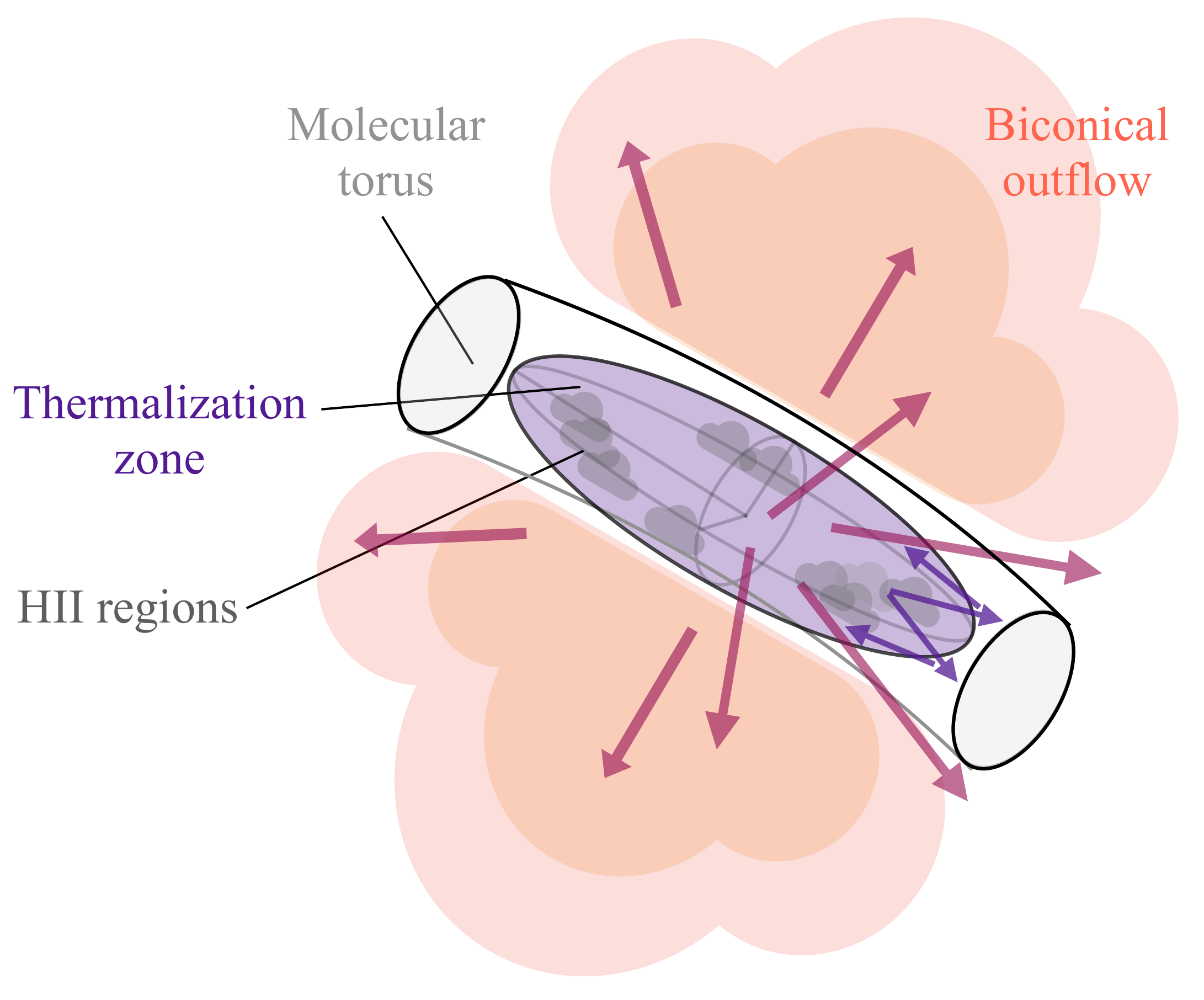}
\caption{\textbf{Geometry of the starburst nucleus of M82.} The hot, Fe~{\sc xxv}-bearing wind fluid is found within an ellipsoidal thermalization zone (purple), where star clusters demarcated by H~{\sc ii} regions (gray clouds) shock-heat the ambient medium. This zone is bounded in the plane of the galaxy by a molecular torus \cite{1987PASJ...39..685N} (white ring, shown as a cross-section) seen in CO emission. This torus nozzles the hot wind fluid as it leaves the thermalization zone (purple vectors), reducing the surface area through which the hot wind can freely stream (red vectors) and producing the biconical outflow.}
\label{cartoon}
\end{figure}

\textit{Chandra} and optical imaging shows that the Fe~{\sc xxv}-bearing gas fills an ellipsoidal volume of dimensions $a \approx 300$~pc, $b \approx 350$~pc, and $c \approx 75$~pc (see Figures~\ref{pointing_fig} and \ref{cartoon} and the Methods). We hereafter refer to this region as the thermalization zone. We find a gas density of $n_{H} \approx 0.7f^{-1/2}$~cm$^{-3}$, a gas mass of $M \approx 6 \times 10^5f^{1/2}$~M$_{\odot}$, and a pressure of $P \approx 3 \times 10^{7}f^{-1/2}$~K~cm$^{-3}$ for a volume filling factor $f$. As hot, unconfined clouds will expand rapidly, we assume $f = 1$. This yields a gas pressure that is consistent with optical spectroscopy of H~{\sc ii} regions associated with star clusters in the thermalization zone \cite{2006MNRAS.370..513S, 2007ApJ...671..358W}.

\begin{figure}[]
\centering
\includegraphics[scale = 0.6]{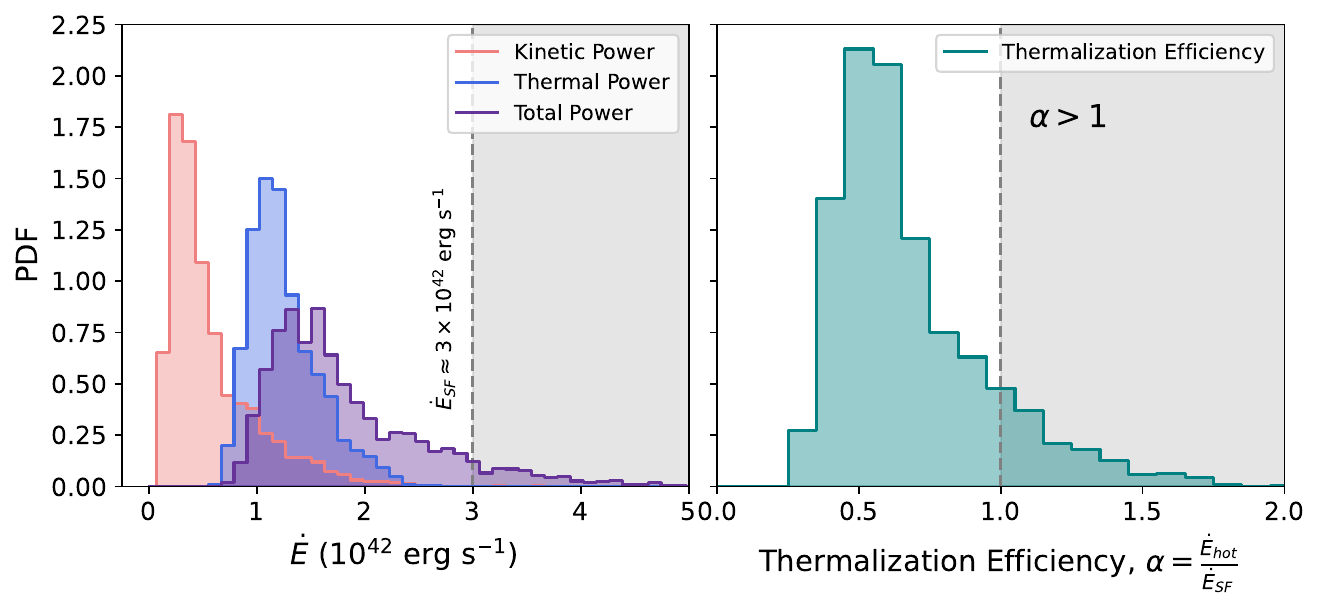}
\caption{\textbf{Power budget and thermalization efficiency of the hot wind.} At left, probability distributions are shown for the thermal (blue), kinetic (red), and total power (purple) carried by the hot wind, as derived from the measurement error on the spectral fit. The total power in the hot wind, $\dot{E}_{\text{hot}}$, yields the thermalization efficiency, or $\alpha = \dot{E}_{\text{hot}}/\dot{E}_{\text{SF}}$, at right. The shaded gray regions indicate where $\dot{E}_{\text{hot}}$ exceeds the $\dot{E}_{\text{SF}} \approx 3 \times 10^{42}$~erg~s$^{-1}$ available from SNe; this is indicative of measurement error and systematic uncertainty on the wind geometry and SN rate.}
\label{therm_eff_fig}
\end{figure}

Based on the same geometry used to derive the nuclear pressure, we can constrain the mass and energy outflow rates of the hot wind. The presence of a biconical outflow and a torus of molecular gas \cite{1987PASJ...39..685N, 2016ApJ...830...72C, 2021ApJ...915L...3K} (outer radius $R \approx 500$~pc, $z \approx 75$~pc) suggest that the outflow is nozzled and occurs over $\approx 45\%$ of the surface bounding the thermalization volume. This yields a mass outflow rate of $\dot{M}_{\text{hot}} \approx 4~M_\odot$~yr$^{-1}$ (see the Methods for a discussion of the wind nozzling and $\dot{M}_{\text{hot}}$). The hot gas mass loss rate to SFR ratio is then $\beta = \dot{M}_{\text{hot}}/\text{SFR} \approx 0.3$ \cite{2018ApJ...864..150V}, consistent with previous estimates of modest $\beta$ values ($\beta \approx 0.2 - 0.5$ \cite{2009ApJ...697.2030S}). This gas has a power of $\dot{E}_{\text{hot}} \approx 2 \times 10^{42}$~erg~s$^{-1}$, of which $\approx 75$\% is thermal (see Figure~\ref{therm_eff_fig}).

\begin{figure}
    \centering
    \includegraphics[width=1.0\linewidth]{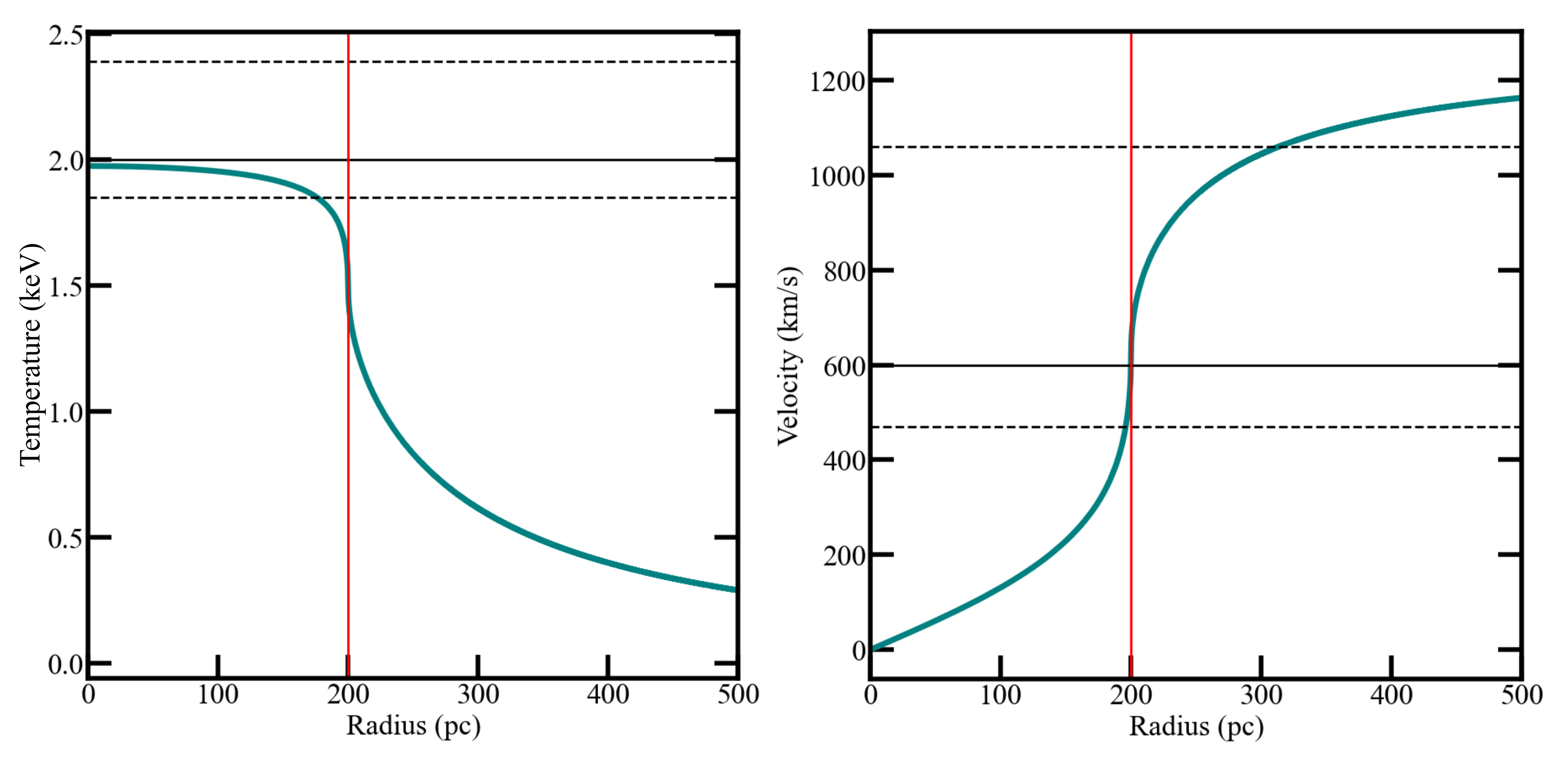}
    \caption{\textbf{A one-dimensional, free wind model of M82.} A free wind model \cite{1985Natur.317...44C} well describes the hot outflow in M82, as shown by temperature (left) and velocity (right) profiles as functions of radius. The black lines show the Resolve measurements and the dashed lines indicate the 90\% confidence intervals. We assume a spherical driving region with a radius of $R_\star = 200$ pc (shown by the red line) and adopt the mass and energy loading inferred from the \textit{XRISM} measurements ($\dot{M} = 4~M_\odot$~yr$^{-1}$, $\dot{E} = 2 \times 10^{42}$~erg~s$^{-1}$).}
    \label{fig:tempmodel}
\end{figure}

A hot ($kT = 2$~keV), high-pressure medium ($P = 3 \times 10^{7}$~K~cm$^{-3}$) expanding at $v_{\text{out}} \approx 600$~km~s$^{-1}$ is generally consistent with expectations from free wind models. We compared the measurements to the one-dimensional analytical free-wind solution derived in \cite{1985Natur.317...44C}, which assumes constant mass and energy loading within a spherical driving region. The solution depends on the energy injection rate, mass injection rate, and driving region radius, as shown in the Methods. Adopting the energy and mass loadings above, the resulting wind has a mean temperature within the driving region ($R_\star = 200$~pc) of $k\bar{T} = 1.9$~keV. Velocities within the driving region are lower than the observed line-of-sight velocity dispersion (see Figure~\ref{fig:tempmodel} and the Methods), although the wind reaches a velocity of $v \approx 660$~km~s$^{-1}$ at the edge of the driving region. These results show that the wind properties in the prototypical starburst are generally well described by a freely expanding wind, although the velocity profile within the driving region may be under-predicted by the classic Chevalier-Clegg model \cite{1985Natur.317...44C}.

Assuming a steady state, the energy outflow rate suggests that at least $\dot{E} \approx 2 \times 10^{42}$~erg~s$^{-1}$ must be continuously supplied by star formation to maintain the hot wind. With knowledge of the SN rate, we can measure what fraction of the SN energy must be thermalized. We define the thermalization efficiency as $\alpha = \dot{E}_{\text{hot}}/\dot{E}_{\text{SF}}$, where $\dot{E}_{\text{SF}}$ is the energy injection rate from SNe. We adopt a star-formation rate of $\text{SFR} = 12.5$~M$_{\odot}$~yr$^{-1}$ \cite{2018ApJ...864..150V} (Methods), leading to an average $\dot{E}_{\text{SF}} \approx 3 \times 10^{42}$~erg~s$^{-1}$. At a rate of $\approx 0.085$~SNe~yr$^{-1}$, SNe occur frequently enough that their energy injection rate is in a quasi-steady state and their average power can be meaningfully compared to the hot wind power. This yields a thermalization efficiency of $\alpha \approx 0.6$, indicating that most SN energy goes into heating the hot gas. This is consistent with previous estimates of moderate to high thermalization efficiency ($0.3 < \alpha < 1$, assuming $\dot{E}_{\text{SF}} \approx 3 \times 10^{42}$~erg~s$^{-1}$ \cite{2009ApJ...697.2030S}). While there are uncertainties on the SFR over the timescale relevant for the wind, independent SN rate measurements from identifying SN remnants yield consistent results (Methods).

The high thermalization efficiency suggests that less than half of the energy from SNe is available to power other feedback modes, such as cosmic rays (expected to be $\sim 10$\% of the SN energy \cite{1987PhR...154....1B}), whose presence is revealed by a synchrotron halo \cite{1991ApJ...369..320S}, and very hot gas at $T > 10^8$~K. The $3\sigma$ upper limit on the Fe~{\sc xxvi}~6.9~keV flux places a limit on the amount of extremely hot, rarefied gas expected from some starburst models \cite{1985Natur.317...44C, 2016MNRAS.455.1830T} of $M < 2.3 \times 10^{5}$~M$_{\odot}$, or $< 35$\% of the mass in the $kT = 2$~keV gas (Supplementary Information). This limit is consistent with the remainder of the SN energy budget, but such gas could not be the dominant phase of the hot wind.

\begin{figure}[h]
\centering
\includegraphics[scale = 0.9]{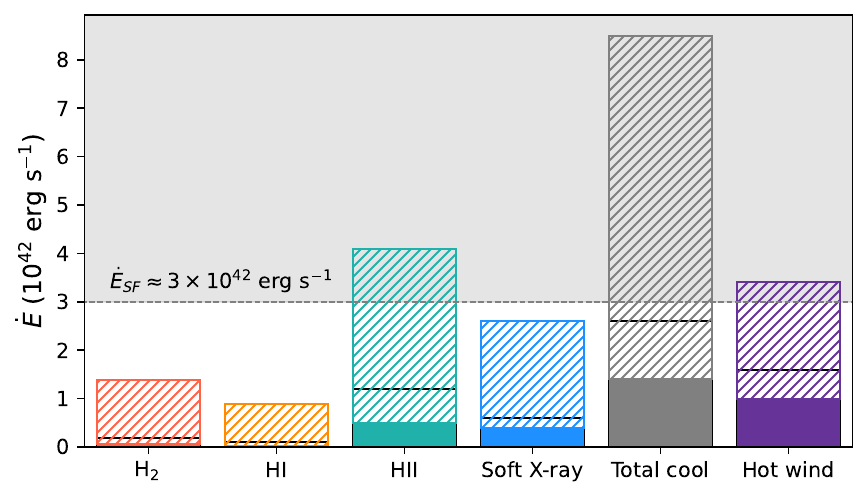}
\caption{\textbf{The multiphase power budget of the wind.} The power available in the hot wind fluid ($\dot{E}_{\text{hot}} = \dot{E}_{\text{th}} + \dot{E}_{\text{kin}}$; purple bar) is comparable to the power requirement of the cooler, multiphase outflow comprised of molecular (H$_{2}$), neutral (H~{\sc i}), warm ionized (H~{\sc ii}), and soft X-ray emitting gas. The total requirement of these cooler phases, $\dot{E}_{\text{cool}}$, is shown by the gray bar. The black lines and hashed regions mark the values and uncertainties given in Table~\ref{cool_wind_tab}. $\dot{E}_{\text{SF}}$ available from SNe is indicated by the horizontal dotted line. For the cool wind, uncertainty in the  wind cone inclination and opening angles contributes to the tail to $\dot{E}_{\text{cool}} >> \dot{E}_{\text{SF}}$.}
\label{cool_wind}
\end{figure}

Assuming a steady-state outflow, we can use existing measurements of molecular, atomic, and ionized gas in the kiloparsec-scale wind to determine whether the Fe~{\sc xxv}-bearing gas has enough power to drive the cooler, multiphase outflow. We consider the total energy outflow rates of the multiphase gas - both thermal and kinetic - and ask whether the hot wind can heat and accelerate the cooler gas. We estimate the power requirement of the multiphase wind, $\dot{E}_{\text{cool}}$, by quantifying the total energy outflow rates of each wind phase from literature measurements (Methods; see also Table~\ref{cool_wind_tab} and Figure~\ref{cool_wind}). The power requirement of the multiphase wind, $\dot{E}_{\text{cool}} = 2.6^{+5.9}_{-1.2} \times 10^{42}$~erg~s$^{-1}$, is dominated by the kinetic term, $\dot{E}_{\text{cool,kin}} \approx 1.6 \times 10^{42}$~erg~s$^{-1}$. Most of the kinetic power is carried by the warm ionized gas ($\dot{E}_{\text{kin}} \approx 1.2^{+2.9}_{-0.7} \times 10^{42}$~erg~s$^{-1}$). The soft X-ray wind ($E \lesssim 4$ keV) accounts for most of the thermal power, with $\dot{E}_{\text{th}} \approx 0.5^{+0.8}_{-0.1} \times 10^{42}$~erg~s$^{-1}$. Within the error, the power of the hot wind, $\dot{E}_{\text{hot}} \approx 1.6^{+1.8}_{-0.6} \times 10^{42}$~erg~s$^{-1}$, is enough to drive the cooler wind (we note that radiative losses, $\dot{E}_{\text{rad}} \approx 2 \times 10^{40}$~erg~s$^{-1}$, are only $\approx 1\%$ of $\dot{E}_{\text{hot}}$). Since the hot wind carries most of the energy from SNe, its energy loading is almost maximally efficient. Other feedback modes, including cosmic rays and hotter gas, may help to launch the wind, but we do not find evidence that they are needed. Similar experiments in a sample of nearby starburst galaxies will provide constraints on stellar feedback prescriptions in hydrodynamical simulations.

The fate of multiphase gas in starburst winds determines the impact of these outflows on their host galaxies and environments. Some of the $\dot{M} \approx 4$~M$_{\odot}$~yr$^{-1}$ of hot gas leaving the thermalization zone will expend its energy by heating and accelerating cooler phases of the wind and will not retain enough energy to reach the intergalactic medium. However, within the uncertainties on the cool and hot wind energy budgets, up to half of the hot wind gas mass, or $\lesssim 2$~M$_{\odot}$~yr$^{-1}$, will retain its energy and escape the galaxy. This suggests that up to $\approx 2 \times 10^{7}$~M$_{\odot}$ have reached the intergalactic medium during the $\approx 10$~Myr since the onset of the recent starbursts \cite{2003ApJ...599..193F}. This gas is metal-enriched from nuclear star formation \cite[e.g.,][]{2024A&A...686A..96F}. Furthermore, the hot wind fluid carries most of the thermal power in the multiphase outflow; the thermal $\dot{E}_{\text{th}} \approx 1.2 \times 10^{42}$~erg~s$^{-1}$ of the $kT = 2$~keV phase exceeds the thermal power in the cool wind by a factor of two (see Table~\ref{cool_wind_tab}). This suggests that up to a third of the total thermal power carried by the cool and hot winds also reaches the intergalactic medium in the hot phase. There is also $T \sim 10^{4} - 10^{7}$~K gas accelerated by the hot wind that is above the galactic escape velocity (see Table~\ref{cool_wind_tab}), further increasing the mass and energy fluxes into the intergalactic medium. M82 is therefore an example of modest heating and chemical enrichment of the intergalactic medium by a low-redshift starburst wind. The era of high-resolution X-ray spectroscopy will enable statistical measurements of the mass, metals, and energy associated with the hottest phases of starburst winds, sharpening our view of the baryon cycle. \\

\textbf{Data and code availability.}
The observational data analyzed in this study will be available in NASA’s High Energy Astrophysics Science Archive Research Center (HEASARC; https://heasarc.gsfc.nasa.gov/) in the summer of 2025.

The codes used for the data reduction (https://heasarc.gsfc.nasa.gov/docs/software/heasoft) and spectral fitting (https://heasarc.gsfc.nasa.gov/xanadu/xspec) are freely available from the HEASARC website.

\clearpage


\begin{table}[h]
\caption{Thermal plasma model of M82 starburst nucleus}%
\begin{tabular}{@{}lccccc@{}}
\toprule
Component & $N_{\text{H}}$ & $kT$ & $v$ & $\sigma$ & Norm, $\mathcal{N}$ \\
   & ($10^{22}$ cm$^{-2}$) & (keV) & (km s$^{-1}$) & (km s$^{-1}$) & ($\propto \int n_{\text{e}} n_{\text{H}} dV$)  \\ 
\midrule
\textbf{Hot wind} & $2.3^{+0.43}_{-0.28}\pm0.7$ & $2.0^{+0.39}_{-0.15}\pm0.2$ & $189^{+250}_{-145}\pm50$ & $595^{+464}_{-128}\pm40$ & $5.6^{+1.2}_{-1.9}\pm1 \times 10^{-3}$ \\
\textbf{Extended Nebula} & $< 0.9$ & $0.72^{+0.10}_{-0.08}\pm0.02$ & $34^{+47}_{-49}\pm5$ & $175^{+86}_{-73}\pm10$ & $2.0^{+0.8}_{-0.2}\pm0.8 \times 10^{-2}$ \\
\botrule
\end{tabular}
\label{bf_model_tab}
\begin{tablenotes}
  \small
  \item The statistical errors are the $90$\% confidence intervals inferred from the MCMC chain, and the systematic errors are determined by adopting different solar abundance tables in \texttt{Xspec}. The model has a $C$-statistic/dof of $13853/14653$.
\end{tablenotes}
\end{table}

\begin{table}[h]
\caption{Power requirement of the cool wind}%
\begin{tabular}{@{}lcccccccc@{}}
\toprule
Phase & $T$ & $v_{\text{out}}$\footnotemark[1] & $\dot{M}$\footnotemark[1] & $\dot{E}_{\text{th}}$ & $\dot{E}_{\text{kin}}$ & $\dot{E}$ & References \\
 & (K) & (km/s) & ($M_{\odot}$/yr) & ($10^{42}$ erg/s) & ($10^{42}$ erg/s) & ($10^{42}$ erg/s) &  \\
\midrule
\textbf{H$_{2}$} & 30 & $175^{+170}_{-60}$\footnotemark[2] & $19^{+18}_{-6}$ & $2.2^{+2.2}_{-0.7} \times 10^{-6}$ & $0.18^{+1.2}_{-0.12}$ & $0.18^{+1.2}_{-0.12}$ & \cite{2002ApJ...580L..21W, 2015ApJ...814...83L}  \\
\textbf{H~{\sc I}} & 400 & $145^{+145}_{-45}$\footnotemark[2] & $17^{+17}_{-6}$ & $5.3^{+5.3}_{-1.8} \times 10^{-5}$ & $0.11^{+0.77}_{-0.08}$ & $0.11^{+0.77}_{-0.08}$ & \cite{Contursi13, 2018ApJ...856...61M} \\
\textbf{H~{\sc II}} & $10^4$ & $660^{+320}_{-160}$\footnotemark[3] & $9^{+4}_{-2}$ & $1.4^{+0.7}_{-0.3} \times 10^{-3}$ & $1.2^{+2.9}_{-0.7}$ & $1.2^{+2.9}_{-0.7}$ & \cite{1998ApJ...493..129S} \\
\textbf{Soft X-ray} & $0.5 - 1 \times 10^7$ & $245^{+350}_{-65}$\footnotemark[4] & $5^{+7}_{-1}$ & $0.5^{+0.8}_{-0.1}$ & $0.09^{+1.26}_{-0.06}$ & $0.6^{+2.0}_{-0.2}$ & \cite{2020ApJ...904..152L}; this work \\
\midrule
\textbf{Total} & & & & & & $2.6^{+5.9}_{-1.2}$ & \\
\botrule
\end{tabular}
\footnotetext[1]{Deprojected average $v_{\text{out}}$ and total $\dot{M}$, as defined through surfaces parallel to the disk at $|z| = 1$ kpc. Uncertainties correspond to $\pm5^{\circ}$ in inclination angle for H$_{2}$, H~{\sc i}, and H~{\sc ii}.} 
\footnotetext[2]{Deprojected $v_{\text{out}}$ for wind cone inclination $i_{\text{w}} = 10^\circ$ (i.e., wind is perpendicular to the disk).}
\footnotetext[3]{Deprojected $v_{\text{out}}$ for $i_{\text{w}} = 15^\circ$ \cite{1998ApJ...493..129S}.}
\footnotetext[4]{We deproject $\sigma$ assuming an opening half-angle of $\theta = 45^{\circ} \pm 30^{\circ}$.}
\label{cool_wind_tab}
\end{table}

\clearpage

\noindent \textbf{{\Large Methods}} \\

\noindent \textbf{\textit{XRISM} observations and data reduction.} \textit{XRISM} observed M82 from 2024 12 May to 14 May (ObsID 300068010) for 239~ks. \textit{XRISM} simultaneously observes with a high resolution spectrometer (Resolve) and a soft X-ray imaging spectrometer (Xtend). This analysis is based on the Resolve data. M82 was located at the Resolve aimpoint (RA = 148.9705$^{\circ}$, dec = +69.6794$^{\circ}$), with a position angle of $\text{PA} = 287^{\circ}$ (see Figure~\ref{pointing_fig}). During the observation, Resolve's aperture gate valve was closed and the filter wheel was set to open position. 

The Resolve data were calibrated using standard \textit{XRISM} pipeline software and the calibration database available in HEASoft v6.34. After calibration, we filtered the data to exclude non-X-ray events with atypical rise times or which were flagged for pixel-to-pixel coincidence (STATUS[4]!=b0). The observation coincided with a geomagnetic storm, so to eliminate periods of high particle background or foreground from solar wind charge exchange, we binned the events in time (using 500~s bins) and flagged periods with background flares. We filtered out one time bin with a significantly elevated count rate of 0.7~ct~s$^{-1}$. After filtering, the exposure time is $t = 207166$~s. We then extracted a spectrum from only high resolution events \cite{2018JATIS...4a1217I}. The spectrum includes events from all pixels except pixel 12 (a calibration pixel not exposed to the sky) and pixel 27, which has anomalous gain behavior. 

To model the spectrum, we generated two response files, a redistribution matrix file (RMF) and ancillary response file (ARF), which encode the effective area and the probability of detecting an incident X-ray of some energy in any detector energy channel. There are several standard RMFs available, of which the ``large'' is most appropriate for this dataset (generated using the \texttt{rslmkrmf} task with \texttt{whichrmf=L} and parameters from \texttt{xa\_rsl\_rmfparam20190101v006.fits}). The effective area depends on the fraction of light from a given source that falls on the $3^{\prime} \times 3^{\prime}$ array and the fraction of events classified as high resolution. In the $E \approx 2 - 10$~keV Resolve bandpass, the M82 nebula is small compared to the $1.2^{\prime}$ point-spread function (Figure~\ref{pointing_fig}) so we assume the source is point-like for ARF construction. We estimate the impact of using a point-source ARF by convolving  \textit{Chandra} images with the \textit{XRISM} PSF, finding that the true extent of the M82 nebula leads to $<$2\% errors in the effective area. The fraction of high resolution events depends on the incident count rate so the effective area is susceptible to errors caused by including anomalous low-resolution secondary events in the total rate\footnote{\url{https://heasarc.gsfc.nasa.gov/docs/xrism/analysis/ttwof/}}. M82 produces a low count rate of $<1$~count~s$^{-1}$ where we expect almost none of these events to be caused by true X-rays, so we excluded them from the effective area calculation. \\

\noindent \textbf{Resolve spectral modeling.} \textit{Chandra} images show that both diffuse gas and X-ray binaries (XRBs) contribute to the Resolve spectrum. \textit{XRISM} does not spatially resolve these components, so we modeled the Resolve spectrum as the sum of emission from plasma in collisional ionization equilibrium (CIE) and continuum emission and a fluorescent Fe~K$\alpha$ (6.4~keV) line from the XRBs. We use the \texttt{Xspec} spectral fitting package with the C-stat fit statistic. We use the Atomic Plasma Emission Code (APEC) \citep{2012ApJ...756..128F} to model the CIE component and a power law to model the emission from the sum of the XRBs. Each component also suffers losses from photoelectric absorption by neutral gas along the line of sight. The basic model is then \texttt{phabs(bapec) + phabs(powerlaw + gauss))}, where \texttt{bapec} includes line-of-sight velocity broadening. We modeled the spectrum over the $E = 1.8 - 10$~keV bandpass where Resolve is well calibrated. Initial modeling showed that the emission lines have measurable velocity broadening and that there are at least two velocity components, as the prominent Fe~{\sc xxv}~6.7~keV line is significantly broader than strong, low energy lines such as S~{\sc xv}~2.4~keV. The simplest model that matches the data is \texttt{phabs$_1$(bapec$_1$ + powerlaw + gauss) $+$ phabs$_2$(bapec$_2$)}, where based on \textit{Chandra} images we expect the hottest component to be absorbed by the same foreground column as the brightest XRBs. 

We obtain confidence intervals on the model parameters using the Goodman-Weare Markov Chain Monte Carlo (MCMC) algorithm in \texttt{Xspec}. We construct MCMC chains with $10^4$ steps taken by 20 walkers. We impose two priors: the velocity centroid of the hot wind is within $200$~km~s$^{-1}$ of the systemic velocity, as expected from a biconical wind, and the metallicity is solar ($Z = Z_{\odot}$) using the solar abundances from \cite{2009ARA&A..47..481A} (see Supplementary Information). 

The best-fit model is shown in Figure~\ref{full_spec}. The model is an excellent match to the data over the full bandpass, but appears to underpredict the red wing of the broad Fe~{\sc xxv}~6.7~keV line. We find cosmetic improvement when fitting the model over a narrow band ($E = 6-9$~keV) around the line, which is achieved via a large redshift ($v - v_{\text{sys}} = +480^{+490}_{-510}$~km~s$^{-1}$) and increased temperature. First, this redshift is not statistically significant; there is no statistical improvement over the broadband fit. Secondly, the redshift is significantly larger than that of S~{\sc xvi}~2.6~keV, which is $v - v_{\text{sys}} = -250^{+270}_{-260}$~km~s$^{-1}$. Since we expect S~{\sc xvi} and Fe~{\sc xxv} to coexist in about the same temperature range, we do not expect to see a large offset between the two. Thirdly, the larger temperature in the narrowband fit requires increased velocity broadening to suppress the Fe~{\sc xxvi}~6.9~keV line, which is not visible in the spectrum. This broadening is again inconsistent with that of S~{\sc xvi}, and the temperature is inconsistent with the S~{\sc xvi}/S~{\sc xv} ratio. We conclude that the broadband model better describes the data, but note that using the narrowband model does not significantly change the inferred energy in the hot phase as temperature and density trade against each other in the models to match the data. We investigated the importance of the non-X-ray background (NXB) and conclude that it has a negligible impact on our analysis (Supplementary Information). \\

\noindent \textbf{\textit{XRISM} best-fit model.} We adopt the 50th percentile MCMC chain parameters as the best-fit model and report 90\% confidence intervals for statistical error bars (see Table~\ref{bf_model_tab}). We also include a systematic error term associated with the adopted solar abundance table (Supplementary Information). The Fe~{\sc xxv}-bearing gas is hot ($kT=2.0^{+0.39}_{-0.15}\pm0.2$~keV) and has a large velocity dispersion ($\sigma = 595^{+464}_{-128}\pm40$~km~s$^{-1}$) after accounting for thermal Doppler broadening. The velocity centroid is consistent with the systemic velocity of M82 ($v - v_{\text{sys}} = -80^{+250}_{-145}\pm50$ km~s$^{-1}$). The foreground absorbing column density of $N_{\text{H}} = 2.3^{+0.43}_{-0.28}\pm0.7 \times 10^{22}$ cm$^{-2}$ is consistent with the large column density observed towards the bright XRBs in the central region \cite{2020ApJ...889...71B}. The best-fit temperature is qualitatively consistent with the emission-line ratios. The non-detection of Fe~{\sc xxvi}~6.9~keV (a $3\sigma$ upper limit of $F(\text{Fe~{\sc xxvi}}) < 3 \times 10^{-6}$ ph~s$^{-1}$~cm$^{-2}$) sets a limit on the temperature of the Fe~{\sc xxv}-bearing gas. The lower limit on the Fe~{\sc xxv}/Fe~{\sc xxvi} flux ratio alone, $F(\text{Fe~{\sc xxv}})/F(\text{Fe~{\sc xxvi}}) > 5.0$, yields a $3\sigma$ upper limit of $kT < 5.5$ keV, or $T < 6 \times 10^{7}$ K. Likewise, the S~{\sc xvi} (2.6~keV)/S~{\sc xv} (2.4~keV) flux ratio limits $kT \gtrsim 1.8$~keV. Since this component is concentrated in the starburst nucleus \cite{2007ApJ...658..258S,2014MNRAS.437L..76L}, we identify it as the ``wind fluid'' from starburst models \cite{1985Natur.317...44C, 2009ApJ...697.2030S}. 

The second thermal plasma component is cooler ($kT = 0.72^{+0.10}_{-0.08}\pm0.02$ keV) and has a lower velocity dispersion ($\sigma = 175^{+86}_{-73}\pm10$ km~s$^{-1}$). It is also less absorbed ($N_{\text{H}} < 0.9 \times 10^{22}$ cm$^{-2}$) and is modestly blueshifted with respect to the galaxy ($v - v_{\text{sys}} = -235^{+47}_{-49}\pm5$ km~s$^{-1}$). Narrow-band \textit{Chandra} images dominated by the strong lines seen with \textit{XRISM},  including Si~{\sc xiii}~1.8~keV, Si~{\sc xiv}~2.0~keV, and S~{\sc xv}~2.4~keV, show that this plasma extends to a characteristic scale of $R \approx 1$~kpc. Based on the spatial extent and velocity offset, we identify this component with the extended hot wind \citep{2020ApJ...904..152L}. \\

\noindent \textbf{Geometry of the Fe~{\sc xxv}-bearing region.} We determined the volume of the Fe~{\sc xxv}-emitting region using a combination of archival \textit{Chandra} narrow-band images (Supplementary Information) and multiwavelength data from the Advanced Camera for Surveys on the \textit{Hubble Space Telescope} (\textit{HST}) and the Near Infrared Camera on \textit{JWST}. We created a narrowband, fluxed Fe~{\sc xxv} image from events in the $E=6.5-6.8$~keV bandpass using the effective area at $E=6.65$~keV. This bandwidth is twice as wide as the 150~eV energy resolution for the ACIS-S3 detector. We created a companion continuum image in the $E=6.8-7.1$~keV bandpass. We used the publicly available, pipeline-processed \textit{HST} (H$\alpha$ and F814W) and \textit{JWST} (2.5$\mu$m) images obtained from the Mikulski Archive for Space Telescopes\footnote{\url{https://mast.stsci.edu/portal/Mashup/Clients/Mast/Portal.html}}.

The hot gas traced by Fe~{\sc xxv} is aligned with the massive star clusters traced by H$\alpha$ and the population of older stars seen at 2.5~$\mu$m, although the starburst ring is not centered on the stellar disks revealed by \textit{JWST} or the \textit{HST} F814W image. This alignment is corroborated by the locations of individual supernova remnants \cite{2010MNRAS.408..607F}. The entire H$\alpha$ complex is surrounded by a molecular ring \cite{1987PASJ...39..685N,2016ApJ...830...72C, 2021ApJ...915L...3K}. Thus, we conclude that the source of Fe~{\sc xxv} is the massive star clusters and that the emission is confined to the nuclear region by the molecular torus, except where it escapes as a wind.

We used least-squares fitting to determine the shape of the star-forming ring, which we then use as a constraint to determine the volume containing the hot gas. We prepared the H$\alpha$ image by clipping at a contour of 1.5~counts~s$^{-1}$ within a circle of radius $R=0.5'$, centered on the nucleus, to isolate the starburst ring. We also masked point sources. We then fit a uniformly dense, optically thin torus model to the patchy surface brightness map. The free parameters are semimajor axis $a$, semiminor axis $b$, height $z$ from the midplane, torus centroid, and position angle (PA). We fixed $z=75$~pc based on the molecular data \cite{2016ApJ...830...72C,2021ApJ...915L...3K}. We fit the data by projecting the 3D model using a viewing angle of 76.5$^{\circ}$ \cite{2005ApJ...628L..33M}, finding $a = 240\pm20$~pc, $b=280\pm40$~pc, and PA$=330\pm4^{\circ}$. 

We then fixed the centroid, PA, and aspect ratio $b/a$ and fitted the model again to the Fe~{\sc xxv} image to determine $a$ for the hot gas. The hot gas does not form a torus, so we smoothed the image with a Gaussian kernel of width $\sigma=5$~pixels and clipped the image at 2$\sigma$ above local background, then fitted an ellipse to the 2$\sigma$ contour. This returns $a=300\pm30$~pc and thus $b=350$~pc. Assuming that $c=z=75$~pc, as we cannot measure this value, this results in a volume $V = 9.7\times 10^{62}$~cm$^{3}$. The size of the ellipsoid indicates that the hot gas is confined by the molecular ring, which nozzles the outflowing gas to produce a biconical outflow. Assuming $z = 75$ pc and $r = 350$ pc as the height and inner radius of the torus, respectively, we estimate that gas escapes from $\approx45$\% of the ellipsoid surface, or $A \approx 3.2 \times 10^{42}$ cm$^{2}$. We adopt this surface area when determining mass and energy outflow rates of the hot wind fluid, as discussed below. \\

\noindent \textbf{Mass and energy outflow rates.} We calculated the mass, $M$, and energy, $E$, of the hot fluid wind and their outflow rates ($\dot{M}$, $\dot{E}$) as follows. The normalization, $\mathcal{N}$, of the \texttt{bapec} model takes the form $\mathcal{N} = \frac{10^{-14}}{4 \pi [D_{\text{A}}(1 + z)]^{2}} \int n_{\text{e}} n_{\text{H}} dV$, where $D_{\text{A}}$ is the angular diameter distance (cm), $n_{\text{e}}$ and $n_{\text{H}}$ are the electron and H densities (cm$^{-3}$), and $dV$ is the volume element (cm$^{3}$). Assuming $D_{\text{A}} = 3.5$ Mpc, $\int dV = 9.7\times 10^{62}$ cm$^{3}$ (see above), and a fully ionized medium with constant density, we estimate $n_{\text{H}} \approx 0.7$ cm$^{-3}$ and $M \approx 6 \pm 1 \times 10^5 M_\odot$ within the thermalization zone. This assumes that half of the Fe~{\sc xxv} luminosity arises within the ellipsoidal volume defined above, as suggested by the \textit{Chandra} image. 

The energy of the hot wind fluid is $E = E_{\text{th}} + E_{\text{kin}} = \frac{3}{2}NkT + \frac{1}{2}Mv_{\text{out}}^{2}$, where $N$ is the total number of particles and $v_{\text{out}}$ is the outflow velocity. We find $E_{\text{th}} \approx 6 \times 10^{54}$~erg and $E_{\text{kin}} \approx 2 \times 10^{54}$~erg, for a total $E \approx 8.3^{+2.8}_{-1.3} \times 10^{54}$~erg. Here we assume $v_{\text{out}} = \sigma$; as this is almost certainly a lower limit to the true outflow velocity (see Main Body), we underestimate the kinetic energy. However, as the energy budget is dominated by the thermal term, the impact on the total budget is likely small.

We estimated the mass and energy outflow rates, $\dot{M} = (M/V) \sigma A$ and $\dot{E} = (E/V) \sigma A$, assuming that the nozzled hot wind flows out at $v_{\text{out}} = \sigma$ through surface area $A$ (see above). This yields $\dot{M} \approx 4^{+2}_{-1}$~M$_{\odot}$~yr$^{-1}$ and $\dot{E} \approx 1.6^{+1.8}_{-0.6} \times 10^{42}$~erg~s$^{-1}$, with $\approx 75$\% of the power carried by the thermal term. As these error estimates include only measurement error and do not account for the assumed geometry, outflow speed, and chemical abundances, we consider the inferred mass and energy and their outflow rates to be approximations. Since we expect $v_{\text{out}} > \sigma$, the true outflow rate is likely higher, although it is constrained by the available supernova energy (Main Body). The outflow rate can only be lower due to statistical error or uncertainty on the wind volume.   \\

\noindent \textbf{Free wind models.} We adopt the free-wind model of \cite{1985Natur.317...44C}, hereafter the CC model. The one-dimensional solution depends on the energy injection rate, $\dot{E}$, the mass injection rate, $\dot{M}$, and the driving region radius, $R_\star$. This yields the solution 
\begin{equation}
    \left( \frac{3\gamma+1/\mathcal{M}^2}{1+3\gamma}\right)^{-\frac{3\gamma+1}{5\gamma+1}} \left( \frac{\gamma-1+2/\mathcal{M}^2}{1+\gamma}\right)^{\frac{\gamma+1}{10\gamma+2}}=\frac{r}{R_\star},
\end{equation}
when $r<R_\star$. When $r>R_\star$, the solution takes the form
\begin{equation}
    \mathcal{M}^{\frac{2}{\gamma-1}}\left( \frac{\gamma-1+2/\mathcal{M}^2}{1+\gamma}\right)^{\frac{\gamma+1}{2\gamma-2}}=\left( \frac{r}{R_\star}\right)^2.
\end{equation}
In these equations, $\gamma$ is the adiabatic index, $\mathcal{M}$ is the Mach number, and $r$ is the radius. We can find the density $\rho$ by integrating these equations, yielding 
\begin{equation}
    \rho = 
    \begin{cases}
    \frac{q_Mr}{3v_{\text{out}}}, & \text{if $r\leq R_\star$}\\
    \frac{q_MR_\star^3}{3v_{\text{out}}r^2}, & \text{if $r > R_\star$}.
    \end{cases}
\end{equation}
Here $q_M =\dot{M}/V_\star$ for the volume of the wind-driving region $V_\star=\frac{4}{3}\pi R_\star^3$. The outflow velocity is then given by
\begin{equation}
    v_{\text{out}}=\mathcal{M}c_s = \mathcal{M}\sqrt{\gamma P/\rho}, 
\end{equation}
where $P$ is the pressure. We adopt $\gamma = \frac{5}{3}$ and a radius of $R_\star = 200$ pc for the wind driving region to reproduce the volume of the Fe~{\sc xxv}-bearing region, as discussed above. Figure \ref{fig:tempmodel} shows the solution for the gas temperature and velocity as functions of $r$, overlaid with the observational constraints from \textit{XRISM}.

As explored in \cite{2009ApJ...697.2030S}, this solution assumes a spherical outflow and is only applicable very close to the wind-driving region, but this is not a limitation for deriving the temperature and velocity profiles and comparing them to the Resolve measurement. The model does not capture some of the small-scale effects that impact gas motion such as turbulence and interactions with the molecular torus (by construction, there is no bulk motion at the center of the wind-driving region). The \textit{XRISM} measurements provide a standard for more complex models that take into account variations in geometry and small-scale effects. \\

\noindent \textbf{Supernova rate in M82 nucleus.} We estimated the SN rate to calculate the efficiency with which SN energy is converted to thermal energy in the wind-driving region, $\alpha = \dot{E}_{\text{hot}}/\dot{E}_{\text{SF}}$. For an outflow rate of $\sigma = 600$~km~s$^{-1}$ (600~pc~Myr$^{-1}$) and a thermalization zone of size $l \approx 300$~pc, hot gas must be replaced within 0.5~Myr. As this is much shorter than O star lifetimes, the progenitors of the SNe that powered the current wind formed at least $5-10$~Myr ago, so we adopted the far-ultraviolet and 24~$\mu$m infrared star formation rate $\text{SFR(FUV + 24~$\mu$m)} \approx 12.5$~M$_{\odot}$~yr$^{-1}$ from \cite{2018ApJ...864..150V}, which is consistent with other literature estimates of the average SFR over the last $\sim 100$ Myr (e.g., \cite{2018ApJ...861...94H, 2018ApJ...861...95H}). This implies a SN rate of $\approx 0.085$~SNe~yr$^{-1}$, assuming a Salpeter initial mass function \cite{1955ApJ...121..161S}.

The recent star-formation history of M82 is bursty, with two large bursts peaking approximately $5$ and $10$ Myr ago \cite{2003ApJ...599..193F}. However, a census of SN remnants in the nucleus of M82 is consistent with the rate inferred from the typical SFR over the past $\sim 100$ Myr. A census of remnants in the X-ray yields a SN rate of $\approx 0.06^{+0.07}_{-0.03}$~SNe~yr$^{-1}$ \cite{2021A&A...652A..18I}, while similar studies in the radio produce rates ranging from $\approx 0.03 - 0.11$~SNe~yr$^{-1}$ (e.g., \cite{1985ApJ...291..693K, 1994MNRAS.266..455M, 1994ApJ...424..114H, 2008MNRAS.391.1384F, 2010MNRAS.408..607F}). As these rates are consistent with our estimated $\approx 0.085$~SNe~yr$^{-1}$, or $\dot{E}_{\text{SF}} \approx 3 \times 10^{42}$~erg~s$^{-1}$, we adopt this value in our assessment of $\alpha$. We caution, however, that there are additional considerations that are difficult to quantify, including the fact that some SNe may be found outside of the thermalization zone. \\

\noindent \textbf{Power requirement of cool wind.} Here we consider whether the hot wind fluid carries sufficient energy to power the cool, multiphase wind, which we define as all wind phases that are cooler than the Fe~{\sc xxv}-bearing gas (i.e., the H$_{2}$, H~{\sc i}, H~{\sc ii}, and soft X-ray outflows). We define the power of the $i$th phase of the wind as the sum of the thermal and kinetic terms,    
\begin{equation}
    \dot{E_{i}} = \dot{E_{i}}_\text{,th} + \dot{E_{i}}_\text{,kin} = \bigg(\frac{3}{2}\dot{N}kT + \frac{1}{2}\dot{M}v^{2}_{\text{out}}\bigg)\bigg\rvert_{z = 1~\text{kpc}},
\end{equation}
where $\dot{N}$, $\dot{M}$, and $v_{\text{out}}$ are the particle flux, mass flux, and outflow velocity, respectively, through surfaces parallel to the disk at heights of $|z| = 1$ kpc from the midplane. We choose this height because it is difficult to measure $v_{\text{out}}$ much closer to the midplane, where the signal from the outflow becomes contaminated by the rotating disk.

We summarize literature values for $v_{\text{out}}$, $\dot{M}$, and the inferred $\dot{E}$ in Table~\ref{cool_wind_tab}. Beginning with the coldest phases, \cite{2015ApJ...814...83L} present gas mass surface densities of H$_{2}$ (traced by CO) and H~{\sc i} (21-cm emission) along the minor axis, which we scale by the deprojected $v_{\text{out}}$ from \cite{2015ApJ...814...83L} and \cite{2018ApJ...856...61M} to determine $\dot{M}$ at $|z| = 1$ kpc. We note that warm molecular gas traced by H$_{2}$ 2.12 $\mu m$ emission contributes negligibly to the energy requirement \cite{2009ApJ...700L.149V}. Similarly, the dust-to-gas mass ratio estimated for the M82 outflow is $\delta \approx 0.005$ \cite{2015ApJ...814...83L}, and thus we neglect the contribution of dust to the energy budget. For the warm ionized outflow, \cite{1998ApJ...493..129S} estimate that the H$\alpha$ filaments have a typical density of $n_{e, \text{f}} \approx 15$ cm$^{-3}$ and a volume filling factor of $\phi \sim 0.1$. We assume that this density declines with distance from the disk with an exponential scale height of $h_{z} \approx 1$ kpc \cite{2015ApJ...814...83L}, $n_{e}(z) = \phi n_{e, \text{f}}~e^{-z/h_{z}}$, or  $n_{e}(z = 1~\text{kpc}) \approx 0.5$ cm$^{-3}$. We adopt the geometric and kinematic models of \cite{1998ApJ...493..129S} (see their Table~3) to infer $v_{\text{out}}$, $\dot{M}$, and $\dot{E}$ in Table~\ref{cool_wind_tab}. 

For the molecular, neutral, and warm ionized phases, uncertainty in the wind cone inclination with respect to the plane of the sky, $i_{\text{w}}$, and therefore in the deprojected outflow velocity, $v_{\text{out}} = v_{\text{obs}}/\sin{i_{\text{w}}}$, dominates the error budget. This is most important for the kinetic power, as $\dot{E}_{\text{kin}} \propto v_{\text{out}}^{3}$ due to the linear dependence of $\dot{M}$ on $v_{\text{out}}$. For the H$_{2}$ and H~{\sc i} winds, we assume that the wind cone is perpendicular to the disk, $i_{\text{w}} = 10^{\circ}$; for the H~{\sc ii} outflow, we adopt $i_{\text{w}} = 15^{\circ}$ measured by \cite{1998ApJ...493..129S}. For each nominal $i_{\text{w}}$, we draw the inclination angle from a Gaussian distribution with a $90$\% confidence interval of $\Delta i_{\text{w}} = \pm 5^{o}$ to obtain an uncertainty on $v_\text{out}$ and $\dot{E}$.

Resolve provides a precise measurement of the line-of-sight velocity dispersion of the $kT = 0.7$~keV wind, $\sigma = 175^{+86}_{-73}$ km s$^{-1}$, within the inner galaxy ($R \lesssim 0.5$ kpc; see Figure~\ref{pointing_fig}). Limits on the velocity dispersion of Mg~{\sc xii}~1.47~keV from the Reflection Grating Spectrometer on \textit{XMM-Newton}, which traces the soft X-ray wind on $R \approx 1$ kpc scales, are consistent with this value ($\sigma < 370$ km s$^{-1}$ \cite{2024ApJ...975..128B}). We therefore assume that $\sigma$ remains constant with distance from the disk and infer $v_{\text{out}}$ for a given wind cone opening half-angle, $\theta$. There is not consensus on $\theta$ in the literature for the M82 wind (e.g., \cite{1998ApJ...493..129S, 2023ApJ...956..142X}), and we therefore adopt a broad Gaussian distribution with $\theta = 45^{\circ} \pm 30^{\circ}$, where the error bar is the $90$\% confidence interval. This yields an estimated $v_\text{out} = 245^{+350}_{-65}$ km s$^{-1}$. We combine this with the density and temperature model of \cite{2020ApJ...904..152L} to determine $\dot{M}$ and $\dot{E}$ for the soft X-ray wind. We caution that the power requirement inferred for this phase carries significant uncertainty due to the unconstrained geometry of the wind. 

As shown in Table~\ref{cool_wind_tab}, the total power requirement of the multiphase wind is $\dot{E}_{\text{cool}} \approx 2.6^{+5.9}_{-1.2} \times 10^{42}$~erg~s$^{-1}$. The power budget is dominated by the thermal power of the soft X-ray wind ($\dot{E}_{\text{th}} = 0.5^{+0.8}_{-0.1} \times 10^{42}$~erg~s$^{-1}$) and the kinetic power of the warm ionized outflow ($\dot{E}_{\text{kin}} = 1.2^{+2.9}_{-0.7} \times 10^{42}$~erg~s$^{-1}$). While the total power estimate has a tail to large $\dot{E}_{\text{cool}}$, this tail exceeds the total power available from SNe, $\dot{E}_{\text{SF}} \approx 3 \times 10^{42}$~erg~s$^{-1}$ (see above). $\dot{E}_{\text{SF}}$ sets a limit on the energy available from any driving mechanism associated with star formation (thermal gas pressure, cosmic rays, radiation, etc.). This suggests that the tail to high $\dot{E}_{\text{cool}}$ for the cool wind is unphysical and can be attributed to uncertainty in the inclination and opening angles of the wind. \\


\clearpage

\begin{thebibliography}{10}
\expandafter\ifx\csname url\endcsname\relax
  \def\url#1{\burl{#1}}\fi
\expandafter\ifx\csname urlprefix\endcsname\relax\def\urlprefix{URL }\fi
\providecommand{\bibinfo}[2]{#2}
\providecommand{\eprint}[2][]{\url{#2}}
\providecommand{\doi}[1]{\url{https://doi.org/#1}}
\bibcommenthead

\bibitem{1978ApJ...219L..23M}
\bibinfo{author}{{McKee}, C.~F.}, \bibinfo{author}{{Cowie}, L.~L.} \& \bibinfo{author}{{Ostriker}, J.~P.}
\newblock \bibinfo{title}{{The acceleration of high-velocity clouds in supernova remnants.}}
\newblock \emph{\bibinfo{journal}{\apjl}} \textbf{\bibinfo{volume}{219}}, \bibinfo{pages}{L23--L28} (\bibinfo{year}{1978}).

\bibitem{1996ApJ...462..651L}
\bibinfo{author}{{Lehnert}, M.~D.} \& \bibinfo{author}{{Heckman}, T.~M.}
\newblock \bibinfo{title}{{Ionized Gas in the Halos of Edge-on Starburst Galaxies: Evidence for Supernova-driven Superwinds}}.
\newblock \emph{\bibinfo{journal}{\apj}} \textbf{\bibinfo{volume}{462}}, \bibinfo{pages}{651} (\bibinfo{year}{1996}).

\bibitem{1990ApJS...74..833H}
\bibinfo{author}{{Heckman}, T.~M.}, \bibinfo{author}{{Armus}, L.} \& \bibinfo{author}{{Miley}, G.~K.}
\newblock \bibinfo{title}{{On the Nature and Implications of Starburst-driven Galactic Superwinds}}.
\newblock \emph{\bibinfo{journal}{\apjs}} \textbf{\bibinfo{volume}{74}}, \bibinfo{pages}{833} (\bibinfo{year}{1990}).

\bibitem{2001ApJ...557..605S}
\bibinfo{author}{{Scannapieco}, E.}, \bibinfo{author}{{Thacker}, R.~J.} \& \bibinfo{author}{{Davis}, M.}
\newblock \bibinfo{title}{{High-Redshift Galaxy Outflows and the Formation of Dwarf Galaxies}}.
\newblock \emph{\bibinfo{journal}{\apj}} \textbf{\bibinfo{volume}{557}}, \bibinfo{pages}{605--615} (\bibinfo{year}{2001}).

\bibitem{2015ARA&A..53...51S}
\bibinfo{author}{{Somerville}, R.~S.} \& \bibinfo{author}{{Dav{\'e}}, R.}
\newblock \bibinfo{title}{{Physical Models of Galaxy Formation in a Cosmological Framework}}.
\newblock \emph{\bibinfo{journal}{\araa}} \textbf{\bibinfo{volume}{53}}, \bibinfo{pages}{51--113} (\bibinfo{year}{2015}).

\bibitem{2023ARA&A..61..131F}
\bibinfo{author}{{Faucher-Gigu{\`e}re}, C.-A.} \& \bibinfo{author}{{Oh}, S.~P.}
\newblock \bibinfo{title}{{Key Physical Processes in the Circumgalactic Medium}}.
\newblock \emph{\bibinfo{journal}{\araa}} \textbf{\bibinfo{volume}{61}}, \bibinfo{pages}{131--195} (\bibinfo{year}{2023}).

\bibitem{2024ARA&A..62..529T}
\bibinfo{author}{{Thompson}, T.~A.} \& \bibinfo{author}{{Heckman}, T.~M.}
\newblock \bibinfo{title}{{Theory and Observation of Winds from Star-Forming Galaxies}}.
\newblock \emph{\bibinfo{journal}{\araa}} \textbf{\bibinfo{volume}{62}}, \bibinfo{pages}{529--591} (\bibinfo{year}{2024}).

\bibitem{1994ApJ...427..628F}
\bibinfo{author}{{Freedman}, W.~L.} \emph{et~al.}
\newblock \bibinfo{title}{{The Hubble Space Telescope Extragalactic Distance Scale Key Project. I. The Discovery of Cepheids and a New Distance to M81}}.
\newblock \emph{\bibinfo{journal}{\apj}} \textbf{\bibinfo{volume}{427}}, \bibinfo{pages}{628} (\bibinfo{year}{1994}).

\bibitem{2009ApJ...697.2030S}
\bibinfo{author}{{Strickland}, D.~K.} \& \bibinfo{author}{{Heckman}, T.~M.}
\newblock \bibinfo{title}{{Supernova Feedback Efficiency and Mass Loading in the Starburst and Galactic Superwind Exemplar M82}}.
\newblock \emph{\bibinfo{journal}{\apj}} \textbf{\bibinfo{volume}{697}}, \bibinfo{pages}{2030--2056} (\bibinfo{year}{2009}).

\bibitem{2002ApJ...580L..21W}
\bibinfo{author}{{Walter}, F.}, \bibinfo{author}{{Weiss}, A.} \& \bibinfo{author}{{Scoville}, N.}
\newblock \bibinfo{title}{{Molecular Gas in M82: Resolving the Outflow and Streamers}}.
\newblock \emph{\bibinfo{journal}{\apjl}} \textbf{\bibinfo{volume}{580}}, \bibinfo{pages}{L21--L25} (\bibinfo{year}{2002}).

\bibitem{2009ApJ...700L.149V}
\bibinfo{author}{{Veilleux}, S.}, \bibinfo{author}{{Rupke}, D. S.~N.} \& \bibinfo{author}{{Swaters}, R.}
\newblock \bibinfo{title}{{Warm Molecular Hydrogen in the Galactic Wind of M82}}.
\newblock \emph{\bibinfo{journal}{\apjl}} \textbf{\bibinfo{volume}{700}}, \bibinfo{pages}{L149--L153} (\bibinfo{year}{2009}).

\bibitem{2015ApJ...814...83L}
\bibinfo{author}{{Leroy}, A.~K.} \emph{et~al.}
\newblock \bibinfo{title}{{The Multi-phase Cold Fountain in M82 Revealed by a Wide, Sensitive Map of the Molecular Interstellar Medium}}.
\newblock \emph{\bibinfo{journal}{\apj}} \textbf{\bibinfo{volume}{814}}, \bibinfo{pages}{83} (\bibinfo{year}{2015}).

\bibitem{2018ApJ...856...61M}
\bibinfo{author}{{Martini}, P.} \emph{et~al.}
\newblock \bibinfo{title}{{H I Kinematics along the Minor Axis of M82}}.
\newblock \emph{\bibinfo{journal}{\apj}} \textbf{\bibinfo{volume}{856}}, \bibinfo{pages}{61} (\bibinfo{year}{2018}).

\bibitem{1987AJ.....93..264M}
\bibinfo{author}{{McCarthy}, P.~J.}, \bibinfo{author}{{van Breugel}, W.} \& \bibinfo{author}{{Heckman}, T.}
\newblock \bibinfo{title}{{Evidence for Large-Scale Winds from Starburst Galaxies. I. The Nature of the Ionized Gas in M82 and NGC 253}}.
\newblock \emph{\bibinfo{journal}{\aj}} \textbf{\bibinfo{volume}{93}}, \bibinfo{pages}{264} (\bibinfo{year}{1987}).

\bibitem{1998ApJ...493..129S}
\bibinfo{author}{{Shopbell}, P.~L.} \& \bibinfo{author}{{Bland-Hawthorn}, J.}
\newblock \bibinfo{title}{{The Asymmetric Wind in M82}}.
\newblock \emph{\bibinfo{journal}{\apj}} \textbf{\bibinfo{volume}{493}}, \bibinfo{pages}{129--153} (\bibinfo{year}{1998}).

\bibitem{2003ApJ...599..193F}
\bibinfo{author}{{F{\"o}rster Schreiber}, N.~M.}, \bibinfo{author}{{Genzel}, R.}, \bibinfo{author}{{Lutz}, D.} \& \bibinfo{author}{{Sternberg}, A.}
\newblock \bibinfo{title}{{The Nature of Starburst Activity in M82}}.
\newblock \emph{\bibinfo{journal}{\apj}} \textbf{\bibinfo{volume}{599}}, \bibinfo{pages}{193--217} (\bibinfo{year}{2003}).

\bibitem{1987PASJ...39..685N}
\bibinfo{author}{{Nakai}, N.} \emph{et~al.}
\newblock \bibinfo{title}{{A nuclear molecular ring and gas outflow in the galaxy M 82.}}
\newblock \emph{\bibinfo{journal}{\pasj}} \textbf{\bibinfo{volume}{39}}, \bibinfo{pages}{685--708} (\bibinfo{year}{1987}).

\bibitem{2001A&A...365..571W}
\bibinfo{author}{{Wei{\ss}}, A.}, \bibinfo{author}{{Neininger}, N.}, \bibinfo{author}{{H{\"u}ttemeister}, S.} \& \bibinfo{author}{{Klein}, U.}
\newblock \bibinfo{title}{{The effect of violent star formation on the state of the molecular gas in M 82}}.
\newblock \emph{\bibinfo{journal}{\aap}} \textbf{\bibinfo{volume}{365}}, \bibinfo{pages}{571--587} (\bibinfo{year}{2001}).

\bibitem{1999ApJ...523..575L}
\bibinfo{author}{{Lehnert}, M.~D.}, \bibinfo{author}{{Heckman}, T.~M.} \& \bibinfo{author}{{Weaver}, K.~A.}
\newblock \bibinfo{title}{{Very Extended X-Ray and H{\ensuremath{\alpha}} Emission in M82: Implications for the Superwind Phenomenon}}.
\newblock \emph{\bibinfo{journal}{\apj}} \textbf{\bibinfo{volume}{523}}, \bibinfo{pages}{575--584} (\bibinfo{year}{1999}).

\bibitem{2018ApJ...864..150V}
\bibinfo{author}{{Vulic}, N.} \emph{et~al.}
\newblock \bibinfo{title}{{Black Holes and Neutron Stars in Nearby Galaxies: Insights from NuSTAR}}.
\newblock \emph{\bibinfo{journal}{\apj}} \textbf{\bibinfo{volume}{864}}, \bibinfo{pages}{150} (\bibinfo{year}{2018}).

\bibitem{2007ApJ...658..258S}
\bibinfo{author}{{Strickland}, D.~K.} \& \bibinfo{author}{{Heckman}, T.~M.}
\newblock \bibinfo{title}{{Iron Line and Diffuse Hard X-Ray Emission from the Starburst Galaxy M82}}.
\newblock \emph{\bibinfo{journal}{\apj}} \textbf{\bibinfo{volume}{658}}, \bibinfo{pages}{258--281} (\bibinfo{year}{2007}).

\bibitem{2021A&A...652A..18I}
\bibinfo{author}{{Iwasawa}, K.}
\newblock \bibinfo{title}{{X-ray supernova remnants in the starburst region of M 82}}.
\newblock \emph{\bibinfo{journal}{\aap}} \textbf{\bibinfo{volume}{652}}, \bibinfo{pages}{A18} (\bibinfo{year}{2021}).

\bibitem{1985Natur.317...44C}
\bibinfo{author}{{Chevalier}, R.~A.} \& \bibinfo{author}{{Clegg}, A.~W.}
\newblock \bibinfo{title}{{Wind from a starburst galaxy nucleus}}.
\newblock \emph{\bibinfo{journal}{\nat}} \textbf{\bibinfo{volume}{317}}, \bibinfo{pages}{44--45} (\bibinfo{year}{1985}).

\bibitem{2020SPIE11444E..22T}
\bibinfo{author}{{Tashiro}, M.} \emph{et~al.}
\newblock \bibinfo{editor}{{den Herder}, J.-W.~A.}, \bibinfo{editor}{{Nikzad}, S.} \& \bibinfo{editor}{{Nakazawa}, K.} (eds) \emph{\bibinfo{title}{{Status of x-ray imaging and spectroscopy mission (XRISM)}}}.
\newblock (eds \bibinfo{editor}{{den Herder}, J.-W.~A.}, \bibinfo{editor}{{Nikzad}, S.} \& \bibinfo{editor}{{Nakazawa}, K.}) \emph{\bibinfo{booktitle}{Space Telescopes and Instrumentation 2020: Ultraviolet to Gamma Ray}}, Vol. \bibinfo{volume}{11444} of \emph{\bibinfo{series}{Society of Photo-Optical Instrumentation Engineers (SPIE) Conference Series}}, \bibinfo{pages}{1144422} (\bibinfo{year}{2020}).

\bibitem{2000Sci...290.1325G}
\bibinfo{author}{{Griffiths}, R.~E.} \emph{et~al.}
\newblock \bibinfo{title}{{Hot Plasma and Black Hole Binaries in Starburst Galaxy M82}}.
\newblock \emph{\bibinfo{journal}{Science}} \textbf{\bibinfo{volume}{290}}, \bibinfo{pages}{1325--1328} (\bibinfo{year}{2000}).

\bibitem{2005ApJ...628L..33M}
\bibinfo{author}{{Mayya}, Y.~D.}, \bibinfo{author}{{Carrasco}, L.} \& \bibinfo{author}{{Luna}, A.}
\newblock \bibinfo{title}{{The Discovery of Spiral Arms in the Starburst Galaxy M82}}.
\newblock \emph{\bibinfo{journal}{\apjl}} \textbf{\bibinfo{volume}{628}}, \bibinfo{pages}{L33--L36} (\bibinfo{year}{2005}).

\bibitem{2014MNRAS.437L..76L}
\bibinfo{author}{{Liu}, J.}, \bibinfo{author}{{Gou}, L.}, \bibinfo{author}{{Yuan}, W.} \& \bibinfo{author}{{Mao}, S.}
\newblock \bibinfo{title}{{Fe K lines in the nuclear region of M82}}.
\newblock \emph{\bibinfo{journal}{\mnras}} \textbf{\bibinfo{volume}{437}}, \bibinfo{pages}{L76--L80} (\bibinfo{year}{2014}).

\bibitem{2006MNRAS.370..513S}
\bibinfo{author}{{Smith}, L.~J.} \emph{et~al.}
\newblock \bibinfo{title}{{HST/STIS optical spectroscopy of five super star clusters in the starburst galaxy M82}}.
\newblock \emph{\bibinfo{journal}{\mnras}} \textbf{\bibinfo{volume}{370}}, \bibinfo{pages}{513--527} (\bibinfo{year}{2006}).

\bibitem{2007ApJ...671..358W}
\bibinfo{author}{{Westmoquette}, M.~S.} \emph{et~al.}
\newblock \bibinfo{title}{{Hubble Space Telescope Space Telescope Imaging Spectrograph Spectroscopy of the Environment in the Starburst Core of M82}}.
\newblock \emph{\bibinfo{journal}{\apj}} \textbf{\bibinfo{volume}{671}}, \bibinfo{pages}{358--373} (\bibinfo{year}{2007}).

\bibitem{2016ApJ...830...72C}
\bibinfo{author}{{Chisholm}, J.} \& \bibinfo{author}{{Matsushita}, S.}
\newblock \bibinfo{title}{{The Molecular Baryon Cycle of M82}}.
\newblock \emph{\bibinfo{journal}{\apj}} \textbf{\bibinfo{volume}{830}}, \bibinfo{pages}{72} (\bibinfo{year}{2016}).

\bibitem{2021ApJ...915L...3K}
\bibinfo{author}{{Krieger}, N.} \emph{et~al.}
\newblock \bibinfo{title}{{NOEMA High-fidelity Imaging of the Molecular Gas in and around M82}}.
\newblock \emph{\bibinfo{journal}{\apjl}} \textbf{\bibinfo{volume}{915}}, \bibinfo{pages}{L3} (\bibinfo{year}{2021}).

\bibitem{1987PhR...154....1B}
\bibinfo{author}{{Blandford}, R.} \& \bibinfo{author}{{Eichler}, D.}
\newblock \bibinfo{title}{{Particle acceleration at astrophysical shocks: A theory of cosmic ray origin}}.
\newblock \emph{\bibinfo{journal}{\physrep}} \textbf{\bibinfo{volume}{154}}, \bibinfo{pages}{1--75} (\bibinfo{year}{1987}).

\bibitem{1991ApJ...369..320S}
\bibinfo{author}{{Seaquist}, E.~R.} \& \bibinfo{author}{{Odegard}, N.}
\newblock \bibinfo{title}{{A Nonthermal Radio Halo Surrounding M82}}.
\newblock \emph{\bibinfo{journal}{\apj}} \textbf{\bibinfo{volume}{369}}, \bibinfo{pages}{320} (\bibinfo{year}{1991}).

\bibitem{2016MNRAS.455.1830T}
\bibinfo{author}{{Thompson}, T.~A.}, \bibinfo{author}{{Quataert}, E.}, \bibinfo{author}{{Zhang}, D.} \& \bibinfo{author}{{Weinberg}, D.~H.}
\newblock \bibinfo{title}{{An origin for multiphase gas in galactic winds and haloes}}.
\newblock \emph{\bibinfo{journal}{\mnras}} \textbf{\bibinfo{volume}{455}}, \bibinfo{pages}{1830--1844} (\bibinfo{year}{2016}).

\bibitem{2024A&A...686A..96F}
\bibinfo{author}{{Fukushima}, K.}, \bibinfo{author}{{Kobayashi}, S.~B.} \& \bibinfo{author}{{Matsushita}, K.}
\newblock \bibinfo{title}{{Revisiting the abundance pattern and charge-exchange emission in the centre of M 82}}.
\newblock \emph{\bibinfo{journal}{\aap}} \textbf{\bibinfo{volume}{686}}, \bibinfo{pages}{A96} (\bibinfo{year}{2024}).

\end{thebibliography}

\clearpage

\begin{thebibliography}{10}
\expandafter\ifx\csname url\endcsname\relax
  \def\url#1{\burl{#1}}\fi
\expandafter\ifx\csname urlprefix\endcsname\relax\def\urlprefix{URL }\fi
\providecommand{\bibinfo}[2]{#2}
\providecommand{\eprint}[2][]{\url{#2}}
\providecommand{\doi}[1]{\url{https://doi.org/#1}}
\bibcommenthead

\bibitem[36]{Contursi13}
\bibinfo{author}{{Contursi}, A.} \emph{et~al.}
\newblock \bibinfo{title}{{Spectroscopic FIR mapping of the disk and galactic wind of M 82 with Herschel-PACS}}.
\newblock \emph{\bibinfo{journal}{\aap}} \textbf{\bibinfo{volume}{549}}, \bibinfo{pages}{A118} (\bibinfo{year}{2013}).

\bibitem[37]{2020ApJ...904..152L}
\bibinfo{author}{{Lopez}, L.~A.}, \bibinfo{author}{{Mathur}, S.}, \bibinfo{author}{{Nguyen}, D.~D.}, \bibinfo{author}{{Thompson}, T.~A.} \& \bibinfo{author}{{Olivier}, G.~M.}
\newblock \bibinfo{title}{{Temperature and Metallicity Gradients in the Hot Gas Outflows of M82}}.
\newblock \emph{\bibinfo{journal}{\apj}} \textbf{\bibinfo{volume}{904}}, \bibinfo{pages}{152} (\bibinfo{year}{2020}).

\bibitem[38]{2018JATIS...4a1217I}
\bibinfo{author}{{Ishisaki}, Y.} \emph{et~al.}
\newblock \bibinfo{title}{{In-flight performance of pulse-processing system of the ASTRO-H/Hitomi soft x-ray spectrometer}}.
\newblock \emph{\bibinfo{journal}{Journal of Astronomical Telescopes, Instruments, and Systems}} \textbf{\bibinfo{volume}{4}}, \bibinfo{pages}{011217} (\bibinfo{year}{2018}).

\bibitem[39]{2012ApJ...756..128F}
\bibinfo{author}{{Foster}, A.~R.}, \bibinfo{author}{{Ji}, L.}, \bibinfo{author}{{Smith}, R.~K.} \& \bibinfo{author}{{Brickhouse}, N.~S.}
\newblock \bibinfo{title}{{Updated Atomic Data and Calculations for X-Ray Spectroscopy}}.
\newblock \emph{\bibinfo{journal}{\apj}} \textbf{\bibinfo{volume}{756}}, \bibinfo{pages}{128} (\bibinfo{year}{2012}).

\bibitem[40]{2009ARA&A..47..481A}
\bibinfo{author}{{Asplund}, M.}, \bibinfo{author}{{Grevesse}, N.}, \bibinfo{author}{{Sauval}, A.~J.} \& \bibinfo{author}{{Scott}, P.}
\newblock \bibinfo{title}{{The Chemical Composition of the Sun}}.
\newblock \emph{\bibinfo{journal}{\araa}} \textbf{\bibinfo{volume}{47}}, \bibinfo{pages}{481--522} (\bibinfo{year}{2009}).

\bibitem[41]{2020ApJ...889...71B}
\bibinfo{author}{{Brightman}, M.} \emph{et~al.}
\newblock \bibinfo{title}{{Spectral Evolution of the Ultraluminous X-Ray Sources M82 X-1 and X-2}}.
\newblock \emph{\bibinfo{journal}{\apj}} \textbf{\bibinfo{volume}{889}}, \bibinfo{pages}{71} (\bibinfo{year}{2020}).

\bibitem[42]{2010MNRAS.408..607F}
\bibinfo{author}{{Fenech}, D.}, \bibinfo{author}{{Beswick}, R.}, \bibinfo{author}{{Muxlow}, T.~W.~B.}, \bibinfo{author}{{Pedlar}, A.} \& \bibinfo{author}{{Argo}, M.~K.}
\newblock \bibinfo{title}{{Wide-field Global VLBI and MERLIN combined monitoring of supernova remnants in M82}}.
\newblock \emph{\bibinfo{journal}{\mnras}} \textbf{\bibinfo{volume}{408}}, \bibinfo{pages}{607--621} (\bibinfo{year}{2010}).

\bibitem[43]{2018ApJ...861...94H}
\bibinfo{author}{{Herrera-Camus}, R.} \emph{et~al.}
\newblock \bibinfo{title}{{SHINING, A Survey of Far-infrared Lines in Nearby Galaxies. I. Survey Description, Observational Trends, and Line Diagnostics}}.
\newblock \emph{\bibinfo{journal}{\apj}} \textbf{\bibinfo{volume}{861}}, \bibinfo{pages}{94} (\bibinfo{year}{2018}).

\bibitem[44]{2018ApJ...861...95H}
\bibinfo{author}{{Herrera-Camus}, R.} \emph{et~al.}
\newblock \bibinfo{title}{{SHINING, A Survey of Far-infrared Lines in Nearby Galaxies. II. Line-deficit Models, AGN Impact, [C II]-SFR Scaling Relations, and Mass-Metallicity Relation in (U)LIRGs}}.
\newblock \emph{\bibinfo{journal}{\apj}} \textbf{\bibinfo{volume}{861}}, \bibinfo{pages}{95} (\bibinfo{year}{2018}).

\bibitem[45]{1955ApJ...121..161S}
\bibinfo{author}{{Salpeter}, E.~E.}
\newblock \bibinfo{title}{{The Luminosity Function and Stellar Evolution.}}
\newblock \emph{\bibinfo{journal}{\apj}} \textbf{\bibinfo{volume}{121}}, \bibinfo{pages}{161} (\bibinfo{year}{1955}).

\bibitem[46]{1985ApJ...291..693K}
\bibinfo{author}{{Kronberg}, P.~P.}, \bibinfo{author}{{Biermann}, P.} \& \bibinfo{author}{{Schwab}, F.~R.}
\newblock \bibinfo{title}{{The nucleus of M 82 at radio and X-ray bands : discovery of a new radio population of supernova candidates.}}
\newblock \emph{\bibinfo{journal}{\apj}} \textbf{\bibinfo{volume}{291}}, \bibinfo{pages}{693--707} (\bibinfo{year}{1985}).

\bibitem[47]{1994MNRAS.266..455M}
\bibinfo{author}{{Muxlow}, T.~W.~B.} \emph{et~al.}
\newblock \bibinfo{title}{{The structure of young supernova remnants in M82.}}
\newblock \emph{\bibinfo{journal}{\mnras}} \textbf{\bibinfo{volume}{266}}, \bibinfo{pages}{455--467} (\bibinfo{year}{1994}).

\bibitem[48]{1994ApJ...424..114H}
\bibinfo{author}{{Huang}, Z.~P.}, \bibinfo{author}{{Thuan}, T.~X.}, \bibinfo{author}{{Chevalier}, R.~A.}, \bibinfo{author}{{Condon}, J.~J.} \& \bibinfo{author}{{Yin}, Q.~F.}
\newblock \bibinfo{title}{{Compact Radio Sources in the Starburst Galaxy M82 and the Sigma -D Relation for Supernova Remnants}}.
\newblock \emph{\bibinfo{journal}{\apj}} \textbf{\bibinfo{volume}{424}}, \bibinfo{pages}{114} (\bibinfo{year}{1994}).

\bibitem[49]{2008MNRAS.391.1384F}
\bibinfo{author}{{Fenech}, D.~M.}, \bibinfo{author}{{Muxlow}, T.~W.~B.}, \bibinfo{author}{{Beswick}, R.~J.}, \bibinfo{author}{{Pedlar}, A.} \& \bibinfo{author}{{Argo}, M.~K.}
\newblock \bibinfo{title}{{Deep MERLIN 5GHz radio imaging of supernova remnants in the M82 starburst}}.
\newblock \emph{\bibinfo{journal}{\mnras}} \textbf{\bibinfo{volume}{391}}, \bibinfo{pages}{1384--1402} (\bibinfo{year}{2008}).

\bibitem[50]{2024ApJ...975..128B}
\bibinfo{author}{{Boettcher}, E.} \& \bibinfo{author}{{Hodges-Kluck}, E.}
\newblock \bibinfo{title}{{Evidence for a Fast Soft X-Ray Wind in M82 from XMM-RGS}}.
\newblock \emph{\bibinfo{journal}{\apj}} \textbf{\bibinfo{volume}{975}}, \bibinfo{pages}{128} (\bibinfo{year}{2024}).

\bibitem[51]{2023ApJ...956..142X}
\bibinfo{author}{{Xu}, X.}, \bibinfo{author}{{Heckman}, T.}, \bibinfo{author}{{Yoshida}, M.}, \bibinfo{author}{{Henry}, A.} \& \bibinfo{author}{{Ohyama}, Y.}
\newblock \bibinfo{title}{{What Are the Radial Distributions of Density, Outflow Rates, and Cloud Structures in the M82 Wind?}}
\newblock \emph{\bibinfo{journal}{\apj}} \textbf{\bibinfo{volume}{956}}, \bibinfo{pages}{142} (\bibinfo{year}{2023}).

\end{thebibliography}

\clearpage

\begin{thebibliography}{10}
\expandafter\ifx\csname url\endcsname\relax
  \def\url#1{\burl{#1}}\fi
\expandafter\ifx\csname urlprefix\endcsname\relax\def\urlprefix{URL }\fi
\providecommand{\bibinfo}[2]{#2}
\providecommand{\eprint}[2][]{\url{#2}}
\providecommand{\doi}[1]{\url{https://doi.org/#1}}
\bibcommenthead

\bibitem[52]{2016JLTP..184..498P}
\bibinfo{author}{{Porter}, F.~S.} \emph{et~al.}
\newblock \bibinfo{title}{{Temporal Gain Correction for X-ray Calorimeter Spectrometers}}.
\newblock \emph{\bibinfo{journal}{Journal of Low Temperature Physics}} \textbf{\bibinfo{volume}{184}}, \bibinfo{pages}{498--504} (\bibinfo{year}{2016}).

\bibitem[53]{Porter2025_submitted}
\bibinfo{author}{{Porter, F.~S. et al.}}
\newblock \bibinfo{title}{{In-flight performance of the XRISM/Resolve detector system}}.
\newblock \emph{\bibinfo{journal}{Submitted to JATIS}} .

\bibitem[54]{Eckart2025_submitted}
\bibinfo{author}{{Eckart, M.~E. et al.}}
\newblock \bibinfo{title}{{Energy gain scale calibration of the XRISM Resolve microcalorimeter spectrometer: ground calibration results and on-orbit comparison}}.
\newblock \emph{\bibinfo{journal}{Submitted to JATIS}} .

\bibitem[55]{Leutenegger2025_submitted}
\bibinfo{author}{{Leutenegger, M.~A. et al.}}
\newblock \bibinfo{title}{{Core line spread function calibration of the XRISM Resolve x-ray calorimeter spectrometer}}.
\newblock \emph{\bibinfo{journal}{Submitted to JATIS}} .

\bibitem[56]{2020ApJ...895...43S}
\bibinfo{author}{{Schneider}, E.~E.}, \bibinfo{author}{{Ostriker}, E.~C.}, \bibinfo{author}{{Robertson}, B.~E.} \& \bibinfo{author}{{Thompson}, T.~A.}
\newblock \bibinfo{title}{{The Physical Nature of Starburst-driven Galactic Outflows}}.
\newblock \emph{\bibinfo{journal}{\apj}} \textbf{\bibinfo{volume}{895}}, \bibinfo{pages}{43} (\bibinfo{year}{2020}).

\bibitem[57]{2001MNRAS.321L..29K}
\bibinfo{author}{{Kaaret}, P.} \emph{et~al.}
\newblock \bibinfo{title}{{Chandra High-Resolution Camera observations of the luminous X-ray source in the starburst galaxy M82}}.
\newblock \emph{\bibinfo{journal}{\mnras}} \textbf{\bibinfo{volume}{321}}, \bibinfo{pages}{L29--L32} (\bibinfo{year}{2001}).

\bibitem[58]{2009ApJ...692..653K}
\bibinfo{author}{{Kaaret}, P.}, \bibinfo{author}{{Feng}, H.} \& \bibinfo{author}{{Gorski}, M.}
\newblock \bibinfo{title}{{A Major X-Ray Outburst From an Ultraluminous X-Ray Source in M82}}.
\newblock \emph{\bibinfo{journal}{\apj}} \textbf{\bibinfo{volume}{692}}, \bibinfo{pages}{653--658} (\bibinfo{year}{2009}).

\bibitem[59]{2014Natur.513...74P}
\bibinfo{author}{{Pasham}, D.~R.}, \bibinfo{author}{{Strohmayer}, T.~E.} \& \bibinfo{author}{{Mushotzky}, R.~F.}
\newblock \bibinfo{title}{{A 400-solar-mass black hole in the galaxy M82}}.
\newblock \emph{\bibinfo{journal}{\nat}} \textbf{\bibinfo{volume}{513}}, \bibinfo{pages}{74--76} (\bibinfo{year}{2014}).

\bibitem[60]{2014Natur.514..202B}
\bibinfo{author}{{Bachetti}, M.} \emph{et~al.}
\newblock \bibinfo{title}{{An ultraluminous X-ray source powered by an accreting neutron star}}.
\newblock \emph{\bibinfo{journal}{\nat}} \textbf{\bibinfo{volume}{514}}, \bibinfo{pages}{202--204} (\bibinfo{year}{2014}).

\bibitem[61]{2000ApJ...542..914W}
\bibinfo{author}{{Wilms}, J.}, \bibinfo{author}{{Allen}, A.} \& \bibinfo{author}{{McCray}, R.}
\newblock \bibinfo{title}{{On the Absorption of X-Rays in the Interstellar Medium}}.
\newblock \emph{\bibinfo{journal}{\apj}} \textbf{\bibinfo{volume}{542}}, \bibinfo{pages}{914--924} (\bibinfo{year}{2000}).

\bibitem[62]{2003ApJ...591.1220L}
\bibinfo{author}{{Lodders}, K.}
\newblock \bibinfo{title}{{Solar System Abundances and Condensation Temperatures of the Elements}}.
\newblock \emph{\bibinfo{journal}{\apj}} \textbf{\bibinfo{volume}{591}}, \bibinfo{pages}{1220--1247} (\bibinfo{year}{2003}).

\bibitem[63]{2008MNRAS.386.1464R}
\bibinfo{author}{{Ranalli}, P.}, \bibinfo{author}{{Comastri}, A.}, \bibinfo{author}{{Origlia}, L.} \& \bibinfo{author}{{Maiolino}, R.}
\newblock \bibinfo{title}{{A deep X-ray observation of M82 with XMM-Newton}}.
\newblock \emph{\bibinfo{journal}{\mnras}} \textbf{\bibinfo{volume}{386}}, \bibinfo{pages}{1464--1480} (\bibinfo{year}{2008}).

\bibitem[64]{2011PASJ...63S.913K}
\bibinfo{author}{{Konami}, S.}, \bibinfo{author}{{Matsushita}, K.}, \bibinfo{author}{{Go Tsuru}, T.}, \bibinfo{author}{{Gandhi}, P.} \& \bibinfo{author}{{Tamagawa}, T.}
\newblock \bibinfo{title}{{Suzaku Metal Abundance Patterns in the Outflow Region of M 82 and the Importance of Charge Exchange}}.
\newblock \emph{\bibinfo{journal}{\pasj}} \textbf{\bibinfo{volume}{63}}, \bibinfo{pages}{S913--S924} (\bibinfo{year}{2011}).

\bibitem[65]{2014ApJ...794...61Z}
\bibinfo{author}{{Zhang}, S.} \emph{et~al.}
\newblock \bibinfo{title}{{Spectral Modeling of the Charge-exchange X-Ray Emission from M82}}.
\newblock \emph{\bibinfo{journal}{\apj}} \textbf{\bibinfo{volume}{794}}, \bibinfo{pages}{61} (\bibinfo{year}{2014}).

\bibitem[66]{2013ARA&A..51..457N}
\bibinfo{author}{{Nomoto}, K.}, \bibinfo{author}{{Kobayashi}, C.} \& \bibinfo{author}{{Tominaga}, N.}
\newblock \bibinfo{title}{{Nucleosynthesis in Stars and the Chemical Enrichment of Galaxies}}.
\newblock \emph{\bibinfo{journal}{\araa}} \textbf{\bibinfo{volume}{51}}, \bibinfo{pages}{457--509} (\bibinfo{year}{2013}).

\bibitem[67]{1999ApJS..123....3L}
\bibinfo{author}{{Leitherer}, C.} \emph{et~al.}
\newblock \bibinfo{title}{{Starburst99: Synthesis Models for Galaxies with Active Star Formation}}.
\newblock \emph{\bibinfo{journal}{\apjs}} \textbf{\bibinfo{volume}{123}}, \bibinfo{pages}{3--40} (\bibinfo{year}{1999}).

\end{thebibliography}

\clearpage


\bmhead{Acknowledgements}

This work was supported by JSPS KAKENHI grant numbers JP22H00158, JP22H01268, JP22K03624, JP23H04899, JP21K13963, JP24K00638, JP24K17105, JP21K13958, JP21H01095, JP23K20850, JP24H00253, JP21K03615, JP24K00677, JP20K14491, JP23H00151, JP19K21884, JP20H01947, JP20KK0071, JP23K20239, JP24K00672, JP24K17104, JP24K17093, JP20K04009, JP21H04493, JP20H01946, JP23K13154, JP19K14762, JP20H05857, and JP23K03459 and NASA grant numbers 80NSSC20K0733, 80NSSC18K0978, 80NSSC20K0883, 80NSSC20K0737, 80NSSC24K0678, 80NSSC18K1684, and 80NNSC22K1922. LC acknowledges support from NSF award 2205918. CD acknowledges support from STFC through grant ST/T000244/1. LG acknowledges financial support from Canadian Space Agency grant 18XARMSTMA. MS acknowledges the support by the RIKEN Pioneering Project Evolution of Matter in the Universe (r-EMU) and Rikkyo University Special Fund for Research (Rikkyo SFR). AT and the present research are in part supported by the Kagoshima University postdoctoral research program (KU-DREAM). SY acknowledges support by the RIKEN SPDR Program. IZ acknowledges partial support from the Alfred P. Sloan Foundation through the Sloan Research Fellowship. This material is based upon work supported by NASA under award number 80GSFC24M0006. Part of this work was performed under the auspices of the U.S. Department of Energy by Lawrence Livermore National Laboratory under Contract DE-AC52-07NA27344. This work was supported by the JSPS Core-to-Core Program, JPJSCCA20220002. The material is based on work supported by the Strategic Research Center of Saitama University. \\

\bmhead{Author Contributions}

E. Hodges-Kluck, I. Mitsuishi, T. Tsuru, and N. Yamasaki conceived of the program and planned the observation. M. Yukita reduced the data, and C. Kilbourne and M. Loewenstein performed data quality checks. E. Boettcher, E. Hodges-Kluck, and M. Yukita performed spectral fitting of the Resolve data and led the analysis of the wind temperature, velocity, mass and energy outflow rates, and thermalization efficiency. G. Grell performed Resolve spectral fitting to place a limit on the very hot wind. E. Hodges-Kluck analyzed archival Chandra data and the coordinated Swift observation, and M. Yukita jointly analyzed the Resolve data and archival NuSTAR data. S. Grayson and E. Scannapieco performed the free-wind modeling. E. Boettcher led the preparation of the manuscript. K. Ampuku, R. Cumbee, A. Foster, Y. Fujita, K. Fukushima, A. Hornschemeier, R. Kelley, S. Kobayashi, and S. Sasamata contributed to discussions in regular meetings of the \textit{XRISM} M82 Target Team. T. Sato and J. Vink served as internal reviewers. The science goals of \textit{XRISM} were discussed and developed over 7 years by the XRISM Science Team, all members of which are authors of this manuscript. All of the instruments were prepared by the joint efforts of the team. The manuscript was subject to an internal, collaboration-wide review process. All authors reviewed and approved the final version of the manuscript. \\

\bmhead{Supplementary Information}
Supplementary Information is available for this paper. \\

\bmhead{Author Information}

The authors declare no competing interests. Correspondence and requests for materials should be addressed to E. Boettcher (eboettch@umd.edu). \\

\newpage

\noindent \textbf{{\Large Supplementary Information}} \\

\noindent \textbf{Electron Loss Continuum and Non-X-ray Background.} The ``large'' Resolve RMF that we used does not include the electron-loss continuum. The energy of X-rays absorbed in the detector is imparted to electrons in the detector, and some of these electrons may scatter out. This means that an incident X-ray of a given energy may trigger an event at any lower energy. The probability of this occurring for any one photon is very small, but using the large RMF when electron loss is important can cause systematic errors at the low energy end of the spectrum when fitting models. Electron loss can be included by creating an ``extra-large'' RMF (using \texttt{rslmkrmf} with the \texttt{whichrmf=X} option), but because this RMF produces a signal in each detector channel, it significantly slows down model fitting. We did not use the extra-large RMF for our models because the difference in model count rates in any resolution element between the large and extra-large RMFs is less than 2\%. 

The spectrum includes events from the non-X-ray background (NXB), which consists both of direct particle events (cosmic-ray electrons depositing energy in the detector that are not rejected by the anti-coincidence veto) and fluorescent X-rays from ionized atoms in the detector housing. The characteristic NXB can be included in a spectrum using an empirical model\footnote{\url{https://heasarc.gsfc.nasa.gov/docs/xrism/analysis/nxb/nxb_spectral_models.html}}, which we used to verify that the NXB is unimportant in a smaller set of model fits (i.e., with less robust error estimation). The resulting parameters for the hot wind are $kT = 1.98^{+0.35}_{-0.16}$~keV, $v = 155^{+270}_{-130}$~km~s$^{-1}$, $\sigma = 574^{+390}_{-140}$~km~s$^{-1}$, and $\mathcal{N} = 5.3^{+1.4}_{-1.5} \times 10^{-3}$, while the warm gas has $kT = 0.73^{+0.10}_{-0.12}$~keV, $v = 54^{+55}_{-40}$~km~s$^{-1}$, $\sigma = 156^{+90}_{-95}$~km~s$^{-1}$, and $\mathcal{N} = 2.3^{+0.6}_{-0.4} \times 10^{-2}$. These are very close to the values that we find from the main MCMC posterior distributions, so we can ignore the NXB. \\

\noindent \textbf{Resolve energy scale reconstruction.} The precise relationship between pulse height and energy of each Resolve pixel typically changes slowly in time in response to several factors relating to the changing thermal environment of the sensors and their electronics.  Sporadic calibration intervals using the $^{55}$Fe sources on the Resolve filter wheel are scheduled for each observation to provide energy-scale fiducials. These intervals are scheduled at the start and end of each observation, before and after the thermal disruption caused by recycling the 50\--mK cooler, and at intervals with respect to slewing and recycling that enable linear interpolation between the measurements \cite{2016JLTP..184..498P, Porter2025_submitted}. Seven such calibration measurements were conducted for this observation. The current recommended estimates of the average systematic energy-scale uncertainty for Hp events are $0.3$~eV for the $E = 5.4 - 9$~keV band and $1$ eV at lower energies \cite{Eckart2025_submitted}. The additional uncertainty arising from the drift correction is estimated from the fitted shift to the Mn K$\alpha$ spectrum extracted from the calibration pixel continuously illuminated by a collimated $^{55}$Fe source, after calibrating it only using the same time intervals used to calibrate the array and excluding those time intervals from the spectrum. For M82 it measured to be $0.24$~eV. When added in quadrature to the systematic uncertainty, the resulting total uncertainty near $6$~keV is $0.38$~eV. At lower energies where the calibration uncertainty is higher, $1$~eV remains the estimated systematic energy-scale uncertainty because the correction error is expected to scale roughly with energy. The systematic uncertainty in the Hp resolution (FWHM) is estimated to be $0.05$, $0.1$, and $0.2$~eV at $2$, $5$, and $8$~keV, respectively \cite{Leutenegger2025_submitted}. \\

\noindent \textbf{\textit{Chandra} data processing.} This analysis used \textit{Chandra} observations of M82 using the ACIS-S detector in \texttt{VFAINT} mode, including obsIDs 10542, 10543, 10544, 10925, and 10545, totaling 450~ks. Each data set was calibrated by running the \texttt{chandra\_repro} script with the \textit{Chandra} Interactive Analysis of Observations (CIAO v4.16) software. This script identifies the appropriate calibration products from the \textit{Chandra} database and applies the spacecraft attitude solution to filter detected events for true X-rays and project X-rays to the correct sky position. We used the CIAO \texttt{wavdetect} algorithm with \texttt{sigthresh}=$10^{-6}$ to identify point sources using a PSF map for each observation created by the CIAO \texttt{mkpsfmap} script. These sources were registered to their positions as reported in the \textit{Chandra} Source Catalog v2.1 using the CIAO tool \texttt{wcs\_match}. Then, we combined the cleaned, astrometrically corrected events files using the \texttt{merge\_obs} script and created fluxed images in narrow bands to isolate light from strong emission lines. The \textit{Chandra} energy resolution changes across its bandpass, so for each strong line we selected events within twice the full width at half maximum energy resolution, or for He-like triplets the same criterion for the width of a triplet convolved with a Gaussian with the appropriate energy resolution. We verified these windows by extracting ACIS-S spectra from the diffuse plasma within the Resolve field of view to confirm that the selected energy bands are dominated by the line emission. \\ 

\textbf{Non-equilibrium plasma.} The He-like triplets S~{\sc xv}~2.4~keV and Fe~{\sc xxv}~6.7~keV are fit well by a CIE model; their internal line ratios do not imply significant charge exchange or density-dependent suppression of the forbidden lines. However, we explored the possibility of non-equilibrium ionization (NEI) for the hotter component because frequent SNe may create non-equilibrium conditions and an ionizing plasma can allow for higher temperatures ($T>10^8$~K) predicted by some wind models (e.g., \cite{2020ApJ...895...43S}) without significant Fe~{\sc xxvi} emission. We rejected an NEI model for several reasons. First, it has additional free parameters compared to a CIE model but is not statistically preferred. Secondly, the best-fit ionization timescale of $\tau_{\text{ion}} \approx 10^{11}(n/$cm$^{-3}$)~s, when combined with the model normalization and geometry of the emitting region described below, requires a plasma age of less than 3000~yr. This is inconsistent with a steady-state assumption, as it requires approximately $2\times10^{5}$~yr for a $T=10^8$~K gas traveling at the sound speed to traverse the Fe~{\sc xxv}-emitting region ($l \approx 300$~pc). Therefore, regardless of where in that volume new plasma is generated, most of the emission should come from material that has reached CIE.\\

\noindent \textbf{\textit{NuSTAR} constraints on XRB spectrum.} The XRB emission in M82 is dominated by two ultraluminous X-ray sources (ULXs), M82 X-1 (an intermediate mass black hole candidate with peak X-ray luminosity $L \approx 9 \times 10^{40}$ erg s$^{-1}$ \cite{2001MNRAS.321L..29K, 2009ApJ...692..653K, 2014Natur.513...74P}) and M82 X-2 (a pulsar with peak $L \approx 2 \times 10^{40}$ erg s$^{-1}$ \cite{2014Natur.514..202B}). These bright sources dominate the spectrum above $E \approx 3$ keV. The ULXs are not spatially resolved from the hot wind by Resolve, with offsets of $\lesssim 15''$ from the center of M82. We model the XRB emission as a powerlaw with spectral index $\Gamma = 1.8_{-0.03}^{+0.05}$ and normalization $\mathcal{N} = 6.6_{-0.4}^{+0.7} \times 10^{-3}$. However, the ULXs exhibit variability in both spectral shape and amplitude, with evidence for a turnover in the spectrum around $6$ keV \cite{2007ApJ...658..258S}. As this turnover occurs close to the Fe He$\alpha$ emission, we investigated whether using a broken powerlaw instead of a single powerlaw model for the XRB continuum changes the inferred properties of the hot wind.

M82 has been repeatedly observed with \textit{NuSTAR} over the past decade. Archival \textit{NuSTAR} observations therefore capture the ULX spectrum in various spectral states. We simultaneously fit the \textit{XRISM} and \textit{NuSTAR} spectra to determine whether varying the ULX continuum model within the range of known spectral states affects the hot wind measurements. We selected archival \textit{NuSTAR} observations that are longer than $t_{\text{exp}} = 5$ ks and point $< 3.5'$ from the center of M82, as of May 2024. A total of $36$ observations were downloaded and reprocessed using the \texttt{nupipeline} task with HEASoft v6.34 and CALDB version 20240826. 

To avoid high background periods, we set the \texttt{saamode} parameter to ``STRICT'' and the \texttt{tentacle} parameter to ``yes''. We extracted source spectra using a circle with a radius of $45''$, centered on the galaxy. The background was selected from a source-free region within the same observations, using a circle with a radius of $60''$. Although the \textit{NuSTAR} detector background is not spatially uniform, the total background contribution is less than $2$\% of the source in the energy range of interest.

We initially examined the \textit{NuSTAR} spectra by fitting an absorbed broken powerlaw model (\texttt{bknpower}) across the energy range from $E = 3 - 12$ keV. All observations revealed a spectral break at approximately $6.5$ keV. The powerlaw slope for the softer (harder) energies ranged from $\Gamma_{1} = 1.4 - 1.8$ ($\Gamma_{2} = 2.2 - 3.8$). Notably, three observations indicated a much steeper slope ($\Gamma_{1} > 2$). Given that simultaneous observations from \textit{Swift} and \textit{XRISM} suggest a powerlaw slope of $\Gamma = 1.8$, the steeper slopes ($\Gamma_{1} > 2$) may not accurately represent the ULXs during the \textit{XRISM} observation. Consequently, we used 33 \textit{NuSTAR} observations to assess the point-source continuum. 

We then conducted a joint analysis of the Resolve and \textit{NuSTAR} spectra for each \textit{NuSTAR} observation, using the model \texttt{phabs$_{1}$(bapec$_{1}$ + bknpower) + phabs$_{2}($bapec$_{2}$)}. All model parameters were tied together among the Resolve and \textit{NuSTAR} spectra, except for the normalization of the broken powerlaw due to the ULX variability. Across the $33$ \textit{NuSTAR} observations, the powerlaw indices varied from $\Gamma_{1} = 1.4 - 1.8$ in the soft energy band and from $\Gamma_{2} = 2 - 3$ in the harder energy band. The break energy ranged from $6.4 - 7.4$ keV. The best-fit model parameters for the hot wind and extended nebula remained consistent with the findings from the Resolve-only spectral analysis. In particular, distributions of the best-fit hot wind parameters from the $33$ joint \textit{XRISM} and \textit{NuSTAR} fits have $kT = 1.9\pm0.3$~keV, $v = 120_{-6}^{+14}$~km~s$^{-1}$, $\sigma = 518_{-35}^{+15}$~km~s$^{-1}$, and $\mathcal{N} = 5.6_{-0.8}^{+0.2} \times 10^{-3}$, where the uncertainties are the $90$\% confidence intervals. As these distributions are narrow and fall within the measurement error of the Resolve-only analysis, the systematic uncertainty in the ULX continuum model is small and does not affect our conclusions about the hot wind. \\

\noindent \textbf{Metallicity and chemical abundance pattern.} There is evidence from optical spectroscopy that H~{\sc ii} regions surrounding star clusters in M82 have metallicities similar to solar (e.g., \cite{2006MNRAS.370..513S}), and we assumed that $Z = Z_{\odot}$ in our spectral fitting. If the true metallicity, $Z'$, departs from solar, then the model normalization scales by a factor of $Z_{\odot}/Z'$, and the inferred density and mass scales as $\sqrt{Z_{\odot}/Z'}$. Therefore, even for a large departure from solar metallicity (i.e., a factor of $\approx$ 2), the corresponding uncertainty on the hot wind mass and energy is comparable to the uncertainty associated with the measurement error and wind geometry.

We also assumed the solar chemical abundance pattern of \cite{2009ARA&A..47..481A} in our fitting. We explored the impact of this assumption on the inferred properties of the hot wind. We first estimated the error due to uncertainty in the solar abundance pattern itself by repeating the spectral fitting for $Z = Z_{\odot}$ and the suite of solar abundance tables available in \texttt{Xspec}, including \texttt{wilm} \cite{2000ApJ...542..914W} and \texttt{lodd} \cite{2003ApJ...591.1220L}. We estimated the systematic error from the spread in best-fit values; these errors are generally small compared to the measurement errors and are shown in Table~\ref{bf_model_tab}. The uncertainties given throughout the paper include the measurement error only.

We also estimated the error due to a possible non-solar abundance pattern. X-ray CCD and grating spectroscopy measurements of the soft X-ray wind in M82 generally suggest solar or super-solar ratios of $\alpha$ elements to O (e.g., Ne/O, Mg/O) and somewhat sub-solar Fe/O \cite{2008MNRAS.386.1464R, 2011PASJ...63S.913K, 2014ApJ...794...61Z, 2020ApJ...904..152L, 2024A&A...686A..96F}. This may indicate preferential enrichment by core collapse supernovae (CCSNe), as may be expected for a supernova-driven wind. However, precise measurements of the abundance pattern are complicated by a number of factors, including the presence of charge exchange emission. We considered, as a limiting case, the hot wind properties inferred assuming chemical enrichment by pure CCSNe ejecta. We adopted the CCSNe abundance table from \cite{2013ARA&A..51..457N}, assuming an upper mass cutoff of $M_{\star} < 25$~M$_{\odot}$ to approximate the abundance ratios of \cite{2024A&A...686A..96F}. The spectral fit with $Z = Z_{\odot}$ and this abundance table yields a consistent temperature and velocity dispersion as the solar abundance case within the error, but the best-fit normalization is lower by almost an order of magnitude, resulting in decreases in $\dot{M}$ and $\dot{E}$ by factors of $4$ and $5$, respectively. However, this is a more extreme abundance pattern than we expect in reality. Mixing occurs between the CCSNe ejecta and the ISM in the disk, as well as between the wind fluid and the complex mixture of CGM, outflowing gas, and infalling material above and below the disk. This mixing likely introduces less chemically enriched gas ($Z < Z_{\odot}$) with an abundance pattern closer to solar. We therefore expect that, if the abundance pattern departs from solar, it is intermediate between the solar and CCSNe patterns and has $Z \lesssim Z_{\odot}$. Both of these factors result in higher inferred $\dot{E}$, and we estimate that the associated uncertainty is approximately a factor of two, comparable to the measurement error. \\

\noindent \textbf{Contribution of stellar winds to driving the outflow.} Our estimate of the energy injection rate from star formation, $\dot{E}_{\text{SF}} \approx 3 \times 10^{42}$~erg~s$^{-1}$, does not include contributions from stellar winds. For continuous star formation, \texttt{Starburst99} models \cite{1999ApJS..123....3L} predict that stellar winds are most important for the first several tens of Myr. We assume that the continuous, ``quiescent'' star formation has been occurring on a longer timescale, and therefore energy injection by SNe dominates over stellar winds. For bursty star formation, \texttt{Starburst99} indicates that the importance of stellar winds is sensitive to the age of the burst. The strongest burst in M82's recent history, which peaked at an SFR~$\approx 160$~M$_{\odot}$~yr$^{-1}$ approximately $10$ Myr ago \cite{2003ApJ...599..193F}, occurred long enough ago that SNe also dominate over stellar winds. A second burst occurred approximately $5$ Myr ago, peaking at an SFR~$\approx 40$~M$_{\odot}$~yr$^{-1}$ \cite{2003ApJ...599..193F}. At the age of this burst, SNe and stellar winds contribute comparably to the stellar feedback. However, as the amplitude of the earlier burst is $\approx 4$ times that of the later burst, we expect that stellar winds contribute only $\approx 10 - 20$\% of the current $\dot{E}_{\text{SF}}$, and we therefore neglect this contribution. \\

\noindent \textbf{Limit on very hot wind from non-detection of Fe~{\sc xxvi}.} Models for winds driven by thermal pressure typically assume that the wind-driving fluid is hotter than what we measured, up to several times $10^8$~K \cite{1985Natur.317...44C}. The Resolve spectrum traces $kT \approx 2$~keV gas from the nucleus of M82, and it also constrains the amount of any hotter, more rarified, and thus nearly invisible component that would produce weak Fe~{\sc xxvi}~6.9~keV emission. To measure upper limits on the total mass and energy associated with such an ``invisible'' wind, we constructed a three-temperature model with a hotter component added to the $kT_{1} = 2$~keV and $kT_{2} = 0.7$~keV components of the fiducial model. We selected a temperature of $kT_{3} = 5.5$ keV for the hotter component, in reference to the upper limit determined from the $F(\text{Fe~{\sc xxv}})/F(\text{Fe~{\sc xxvi}})$ flux ratio (Methods). The result is only modestly sensitive to the assumed temperature, as the normalization determined from the upper limit on the Fe~{\sc xxvi} flux declines by only a factor of two (i.e., $n_{\text{H}}$ declines by a factor of $\sqrt{2}$) as the temperature rises from $kT_{3} = 5.5$~keV to $kT_{3} = 10$ keV. Adding a third temperature component does not statistically improve the fit to the spectrum and returns a minimal normalization of $\mathcal{N}_{3}/\mathcal{N}_{1} \approx 0.03$ for the very hot component. 

Assuming that the very hot wind fluid occupies the same volume as the $kT_{1} = 2$~keV component, we calculated $3\sigma$ upper limits for the corresponding gas mass of $M < 2.3 \times 10^{5}$~M$_\odot$ and total energy of $E < 6.5 \times 10^{54}$ erg, which represent $< 35$\% and $< 80$\% of the contributions from the $kT_{1} = 2$~keV gas, respectively. The maximum power carried by this hotter component is $\dot{E} < 1.3 \times 10^{42}$ erg~s$^{-1}$. This implies a total energy outflow rate of $\dot{E} = \dot{E}_{1} + \dot{E}_{3} \lesssim 2.9 \times 10^{42}$ erg~s$^{-1}$, comparable to the total power available from SNe ($\dot{E}_{\text{SF}} \approx 3 \times 10^{42}$ erg~s$^{-1}$; see Methods). In summary, a very hot, undetected wind can at most modestly increase the total hot gas mass by $\lesssim 35$\% and does not dominate the outflow. \\


\end{document}